\documentclass[12pt]{article}
\usepackage[margin=1.3in]{geometry}

\usepackage{setspace}
\usepackage{authblk}
\usepackage{amsthm,amsmath,amssymb}
\usepackage{cases}
\usepackage{dsfont}
\usepackage{multirow}
\usepackage{float}
\usepackage{graphicx}
\usepackage{caption}
\usepackage{amsthm,amsmath,amssymb}
\usepackage{cases}
\usepackage{dsfont}

\numberwithin{equation}{section}

\def\subst{\substack{d\\=}}

\newtheorem{remark}{Remark}[section]

\begin{document}

\date{}
\title{Pivotal Quantities with Arbitrary Small Skewness }

\author{Masoud M. Nasari\thanks{mmnasari@math.carleton.ca}\\
\small{School of Mathematics and Statistics of Carleton University}\\\small{ Ottawa, ON, Canada} }

\maketitle

\begin{abstract}
\noindent
In this paper we present  randomization methods  to  enhance the accuracy of the central limit theorem (CLT) based inferences about  the population mean $\mu$. We introduce a broad class of randomized versions of the Student $t$-statistic,  the  classical pivot for $\mu$, that continue to possess  the pivotal property for $\mu$ and their  skewness  can be made arbitrarily small, for each fixed sample size $n$.
Consequently, these randomized pivots  admit CLTs with smaller errors.
The  randomization framework  in this paper  also provides   an explicit relation between the precision  of the CLTs for the randomized pivots   and the volume  of their associated  confidence regions for the mean for both univariate and multivariate data.  This  property  allows regulating  the trade-off between the accuracy and the volume of the randomized  confidence regions  discussed  in this paper.

\end{abstract}


\section{Introduction}\label{introduction}
The CLT  is  an essential    tool  for  inferring on parameters of  interest  in a nonparametric framework. The strength of the CLT stems from the fact that, as the sample size increases, the usually  unknown  sampling distribution of a pivot,  a function of the data and an  associated parameter,    approaches the standard normal distribution. This, in turn,  validates  approximating   the percentiles of the sampling distribution of the pivot by those  of the  normal distribution, in both univariate and multivariate cases.

\par
The CLT is an approximation method whose validity relies on large enough  samples. In other words,  the larger the sample size is, the more accurate the inference, about the parameter of interest, based on the  CLT will be. The accuracy  of the CLT can be evaluated  in a number ways. Measuring the distance between the sampling distribution of the pivot and the standard normal distribution is the common feature of   these  methods.   Naturally, the latter  distance is a measure of  the error of the CLT. The most well known methods of evaluating  the CLT's error  are   Berry-Ess\'{e}en inequalities and  Edgeworth expansions. These  methods have been extensively studied   in the literature and many contributions have been made to  the area (cf., for example, Barndorff-Nielsen and Cox  \cite{Barndorff-Nielsen and Cox}, Bentkus \emph{et al.} \cite{Bentkus et al}, Bentkus and G\"{o}tze  \cite{Bentkus and Gotze},  Bhattacharya  and Rao   \cite{Bhattacharya  and Rao},
DasGupta  \cite{DasGupta}, Hall  \cite{Hall}, Petrov  \cite{Petrov}, Senatov  \cite{Senatov}, Shao  \cite{Shao} and  Shorack  \cite{Shorack}.

\par
Despite their   differences, the  Berry-Ess\'{e}en inequality  and the  Edgeworth expansion, when the data have a finite third moment, agree on concluding that, usually, the CLT is in error by a term of order  $O(1/\sqrt{n})$, as $n \rightarrow +\infty$, where $n$ is the sample size. In the literature, the latter  asymptotic conclusion is referred to as the first order accuracy or efficiency of the  CLT.

\par
Achieving  more accurate  CLT  based inferences requires alternative  methods of extracting more information, about the parameter of interest, from a given sample that may not be particularly large.

\par
In this paper we introduce a  method to significantly enhance the accuracy of confidence regions for  the population mean via creating  new pivots  for it based on a given set of data.
More precisely, by employing  appropriately chosen  random weights, we construct  new   randomized pivots for the mean. These randomized pivots are more symmetrical than their  classical counterpart the Student $t$-statistic and, consequently,  they admit  CLTs with smaller errors for both univariate and multivariate data.  In fact, by choosing the random weights appropriately, we will  see  that the CLTs for  the  introduced randomized pivots, under some conventional  conditions,   can    already be second order accurate  (see Sections \ref{Error of Convergence} and \ref{Multivariate Pivots}).

\par
The randomization framework in this paper can  be viewed not only as   an alternative to the inferences based on the  classical CLT,   but also  to   the  bootstrap. The bootstrap, introduced by Efron  \cite{Efron},  is a method  that also tends to   increase   the accuracy of CLT based  inferences (cf., e.g., Hall  \cite{Hall} and Singh  \cite{Singh}). The bootstrap  relies   on repeatedly re-sampling from a given  data set (see, for example,   
Efron and Tibshirani  \cite{Efron  and Tibshirani}).

\par
The  methodology  introduced  in this paper, on the other hand,  reduces  the error of   the  CLT  in  a customary fashion, in  both univariate and multivariate cases,      and it  does not require   re-sampling from the given  data (see Remark \ref{Comparison to the bootstrap} below for a brief comparison between the randomization approach of this paper and  the bootstrap).

\par
For confidence regions   based on CLTs to capture a parameter of interest,   in addition to the accuracy,  it is  desirable   to also   address their  volume.

\par
In this  paper we also address the  volume  of the resulting  confidence regions   based on our   randomized pivotal quantities in both univariate and multivariate cases. In the randomization framework of this paper, and in view of the CLTs  for the   randomized pivots introduced in it, studying the volume   of the resulting randomized confidence regions for the mean   is  rather  straightforward.  This, in turn,  enables   one to  easily trace the effect of the reduction in the error, i.e., the higher accuracy,  on the   volume  of the resulting confidence regions. As a result,  one will be able to   regulate the trade-off  between the precision   and the volume  of the randomized  confidence regions (see Section \ref{Length of the Confidence Intervals for mu}, Section \ref{Multivariate Pivots} and Appendix I).

 \par
The rest of this paper is organized as follows. In Section \ref{Main Results} we introduce the new  randomized  pivots for the mean of  univariate data.  In Section \ref{Error of Convergence} we use    Edgeworth expansions    to explain  how the randomization techniques introduced in Section \ref{Main Results} result in a higher accuracy of the CLT.  In Section  \ref{Length of the Confidence Intervals for mu}, for univariate data, we investigate  the length of the confidence intervals that result from the use of the randomized pivots introduced in Section \ref{Main Results}.
Extensions of the randomization techniques of Section \ref{Main Results} to classes of   triangular   random weights are presented in   Section \ref{Multinomially Weighted Pivots}. Generalization of the results   in Sections \ref{Main Results} and \ref{Edgeworth exapnsions} to vector valued data are presented in Section \ref{Multivariate Pivots}.

\section{Randomized pivots with higher accuracy }\label{Main Results}
Let $X,X_1,\ldots,X_n$, $n\geq 1$, be  i.i.d. random variables with  $E_{X}|X_1|^3<+\infty$,   $\mu:=E_X (X_1)$ and   $\sigma^2_X:=Var_{X}(X_1)>0$. The Student $t$-statistic,  the classical  pivot for $\mu$,   is defined as:

\begin{equation}\label{eq 1}
t_n := \sum_{i=1}^n (X_i -\mu)\big/ (S_n \sqrt{n}),
\end{equation}
where $S_{n}^{2}= \sum_{i=1}^n (X_i-\bar{X}_n)^2/n$ and $\bar{X}_n$ are  the sample variance and the sample mean, respectively. Under the assumption   $E_X |X_1|^3<+\infty$, the Berry-Ess\'{e}en inequality and the Edgeworth expansion unanimously   assert   that,  without restricting the class of distributions of the data,   $t_n$ converges  in distribution  to standard normal at the rate $O(1/\sqrt{n})$, i.e., the CLT for $t_n$ is first order accurate.  We are now to improve upon the  accuracy of $t_n$ by  using   a broad class   of random weights. The  improvement will result from replacing the pivot $t_n$ by  randomized versions of it that continue to serve as pivots for   $\mu$.

\par
We now define the aforementioned randomized pivots for $\mu$, as follows:

\begin{equation}\label{eq 2}
g^{w}_n (\theta):= \sum_{i=1}^n (w_{i}-\theta)(X_{i}-\mu)\big/(S_n \sqrt{n E_{w}(w_{1}-\theta)^2} ),
\end{equation}
where   $w$'s are some  random weights  and    $\theta$, to which we refer as the window,  is a real valued constant. The weights $w$'s  and the window  constant $\theta$ are to be  chosen  according to either one of  the following two scenarios, namely,  Method I and  Method II. 
\\
\subsection*{\emph{Method I}:  Non-centered weights}
To construct the  randomized pivot $g^{w}_n (\theta)$  in this scenario,   we let the weights  $ w_1,\ldots,w_n$   be    a \emph{random sample}  with  $E_{w}\big| w_{1} \big|^3 <+\infty$. Moreover, these  weights     should  be  \emph{independent} from the  data $X_1,\ldots,X_n$.  
The window constant  $\theta$,  should be chosen in such a way  that it satisfies the following two  properties: \\ \\
(i) $\theta\neq E_{w}(w_1)$,
\\ \\
(ii) $\textrm{SRF}^{w}(\theta):=\displaystyle {  E_w(w_{1}-\theta)^3 \big/ \big(E_w (w_{1}-\theta)^2\big)^{3/2}  =\delta}$,
\\\\
where $\delta$ is a given number such that $|\delta|$ can be arbitrary small or zero.
\\
\par
The notation $\textrm{SRF}$  is an abbreviation for  \textbf{S}kewness \textbf{R}educing  \textbf{F}actor (see (\ref{eq 2''}) below for a justification for  this notation).

\begin{remark}
The weights $w$'s in Method I  can be  generated,  independently from the data, using some statistical software. This remark is applicable to all randomized pivots discussed in this paper.
\end{remark}
\noindent
\textbf{Discussion of  Method I: When the weights  have a skewed distribution}
\\
In terms of the error of the CLT, an ideal realization  of  condition (ii) of  Method I could be  when the weights  $w$'s have  a skewed distribution and  the window constant $\theta$ is a real root for the cubic equation  $E_w(w_{1}-\theta)^3=0$, i.e., when $\delta=0$.
Condition (ii) of Method I is so that it also allows  generating  the $w$'s  from  skewed distributions and  finding  a window constant $\theta$ such that   $\theta\neq E_{w}(w_1)$ and   $\textrm{SRF}^{w}(\theta)$ is close  enough  to zero. Hence, when $\delta\neq 0$, but $|\delta|$ is chosen to be small, then  $\theta$ does not necessarily  have to be a  root of the equation  $E_w(w_{1}-\theta)^3=0$.

\par
As it can be inferred from the results  in Section \ref{Error of Convergence}, the closer the value of the $\textrm{SRF}^{w}(\theta)$  is to zero, the smaller the error of the CLT for $g^{w}_n(\theta)$, as in
(\ref{eq 2}), will be.
\\
\\
\textbf{Discussion of Method I: When the weights  are symmetrical about their mean}
\\
When the $w$'s are generated  from a distribution that is symmetrical about its mean,   in view of Method I, a refinement can be achieved by taking the window constant $\theta$  to be   close to $E_w (w_1)$ but not equal to it. This choice of $\theta$ will result in $\textrm{SRF}^{w}(\theta)$ that are  not exactly zero,  but can be   arbitrarily  close to it.

\begin{figure}[htb!]
\vspace{-2 cm}
\includegraphics[width=14 cm]{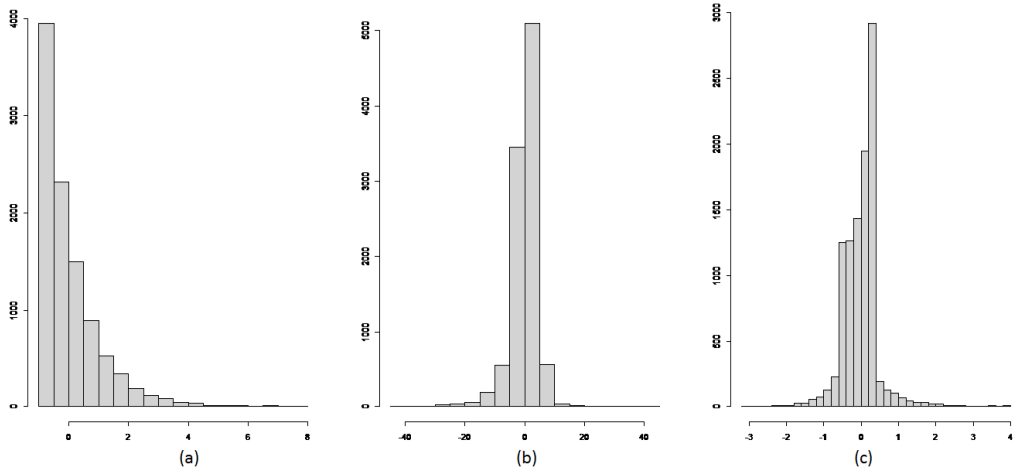}
\vspace{-2  cm}
\centering
\caption{\footnotesize(Illustration of the effect of Method I on univariate data)
\\
Panel (a) is the frequency histogram of the original centered data
$(X_i-1)$, $1\leq i \leq 10000$, where $X_i$'s are i.i.d \textrm{Exponential}(1) with empirical Pearson's measure of skewness equal to 1.98.
 Panel (b) is the frequency histogram of the randomized data   $(w_i-9.3)(X_i-1)$, where the weights $w_i$'s are i.i.d. $\chi^2(7)$,   $\theta=9.3$ and SRF$^{w}(9.3)\approx - 0.622$, with empirical Pearson's measure of skewness equal to $- 1.29$.
 Panel (c) is the frequency
 histogram of the randomized data   $(w_i-0.58)(X_i-1)$, where the weights $w_i$'s are i.i.d. Bernoulli(1/3),   $\theta=0.58$ and $\textrm{SRF}^{w}(0.58)\approx - 0.7$, with empirical Pearson's  measure of skewness equal to $- 1.34$.
}
\end{figure}

Method II that follows can also  be used  to  construct a  more accurate randomized pivot   $g_{n}^{w}(\theta)$, as defined in (\ref{eq 2}),  via generating  the random weights from some symmetrical distributions.

\subsection*{\emph{Method II}: Symmetrical and centered weights}
In this scenario,   we let the weights    $w_1,\ldots,w_n$   be a \emph{random sample} with  a symmetrical (about its  mean) distribution and     $E_{w}| w_{1} |^3 <+\infty$. Moreover, we assume that the weights  are  \emph{independent} from the data  $X_1,\ldots,X_n$ and we take the window constant $\theta$  to be equal to the mean of the random weights, i.e.,  $\theta=E_w (w_1)$.

\par
Taking $\theta=E_{w}(w_1)$ together with  the symmetry of the distribution of the  weights, imply that,  in the scenario of  Method II, we have  $\textrm{SRF}^{w}(\theta)=0$, where $\textrm{SRF}^{w}(\theta)$ is as defined in (ii) of Method I.

\subsection*{Comparing Method I to Method II}
Using the randomized pivot  $g_{n}^{w}(\theta)$, as in (\ref{eq 2}),  and generating  its associated  random weights $w$'s  according to either Method I or Method II, can result in a significant refinement in inferring about  $\mu$. The reason for this   claim  is given in  Section \ref{Error of Convergence}.

\par
In spite of the higher accuracy of $g^{w}_{n}(\theta)$,    provided by  both Method I and Method II, we emphasize that the former  is  more  desirable.   This is so since, in both univariate and multivariate cases,  Method I  yields randomized  confidence regions  for $\mu$ whose volumes   shrink to zero as the sample size $n$ increases to infinity (see  (\ref{eq 9'}) and Appendix I). Method II, on the other hand, fails to yield  shrinking  confidence regions.  In fact, choosing the weights $w$'s for the pivot $g_{n}^{w}(\theta)$ under the scenario of Method II,  yields   confidence regions  for $\mu$ whose volumes, as $n\rightarrow +\infty$,  approach   a limiting distribution rather than vanishing (see (\ref{eq 10}) and Table 3 below).

\begin{remark}\label{Remark 1}
The term $n E_{w}(w_{1}-\theta)^2$ in the denominator of $g^{w}_n(\theta)$, as in (\ref{eq 2}), under both  Methods I and  II,  can,  equivalently,  be replaced by $\sum_{j=1}^n (w_{j} -\theta)^2$.
\end{remark}

\par
In the  above description of different weights in Methods  I and II,  we  excluded  the  case when the weights $w$'s have a  skewed  distribution and $\theta= E_{w} (w_1)$. This case was omitted   since, in general, it does not necessarily provide a refinement in the CLT for the resulting randomized pivot $g_{n}^{w}(\theta)$,  nor  does it result in confidence regions whose volumes shrink to zero, as the sample size increases.

\section{Error of the CLT for $g_{n}^{w}(\theta)$ under Methods I and II}\label{Error of Convergence}
The main idea behind Methods I and II is to transform    the classical  pivot  $t_n$, as in (\ref{eq 1}), to  $g_{n}^{w}(\theta)$, as in (\ref{eq 2}),  that has a  smaller   \emph{skewness}. To further develop the idea, we first note that $g_{n}^{w}(\theta)$ is governed by the joint distribution of the data $X$ and the weights $w$'s. In view of this observation, we let $P_{X,w}$  stand  for the  joint distribution of the data  and the wights, and we  represent  its associated mean by $E_{X,w}$. Recalling now that in both Method I and Method II the weights are independent from the data, we conclude that   $P_{X,w}=P_{X}.P_{w}$  and, consequently, $E_{X,w}=E_{X}.E_{w}$.
\\

Now observe that
\begin{equation}\nonumber
E_{X,w} \big(  (w_1-\theta)(X_1-\mu) \big)= E_{X} (X_1-\mu) E_{w} \big(w_1-\theta \big)=0.E_{w} \big(w_1-\theta \big)=0.
\end{equation}
In view of the preceding observation, we now obtain    the skewness of the random variables $(X-\mu)(w-\theta)$, under both Methods I and II,  as follows:

\begin{eqnarray}
\textrm{skewness
\ of} \ (X-\mu)(w-\theta)&=& \frac{E_{X,w} \big( (X_1-\mu)(w_1-\theta) \big)^3}{\big\{ E_{X,w} \big( (X_1-\mu)(w_1-\theta) \big)^2  \big\}^{3/2}  }\nonumber\\
&=& \Big( \frac{E_{w} (w_1-\theta)^3}{  \big\{ E_{w}(w_1-\theta)^2 \big\}^{3/2}  } \Big)  \Big( \frac{E_{X}(X_1-\mu)^3}{\sigma^{3}_X} \Big)\nonumber\\
&=& \ \textrm{SRF}^{w}(\theta) \big(\frac{E_{X}(X_1-\mu)^3}{\sigma^{3}_X}\big) . \label{eq 2''}
\end{eqnarray}
The second term of the product on  the r.h.s. of (\ref{eq 2''}), i.e., ${E_{X}(X_1-\mu)^3}\big/{\sigma^{3}_{X}}$, is the skewness of the original data. The closer it is to zero the nearer the sampling distribution of  $t_n$,   as defined in (\ref{eq 1}), will be  to the standard normal. However,  one usually has no control over  the skewness of the original data. The idea in Methods I and II is to incorporate  the  random weights $w$'s  and  to  appropriately choose  a window constant   $\theta$
 in such a way  that      $|\textrm{SRF}^{w}(\theta)|
 $ is arbitrarily small. This, in view of (\ref{eq 2''}),  will result in smaller  skewness of the random variables $(X-\mu)(w-\theta)$ (see also Appendix II for the effect of the skewness reduction methods on vector valued data). The latter property,   in turn,  under appropriate conditions,  can  result in the second order accuracy  of  the CLTs for $g_{n}^{w}(\theta)$, as defined in (\ref{eq 2}),  under  both Methods I and II. The accuracy of $g_{n}^{w}(\theta)$ is  to be discussed later in this section in the  univariate case and, in Section \ref{Multivariate Pivots} in the multivariate case.

\par
In view of (\ref{eq 2''}),   it is  now   easy   to appreciate that when  $\theta$ is chosen in such a way that   $\textrm{SRF}^{w}(\theta)=0$, then the skewness of $(X-\mu)(w-\theta)$  will be  exactly zero.  The latter case  can happen under Method I when  the distribution of the $w$'s is  skewed and the cubic equation  $E_{w} (w-\theta)^3=0$  has  at least one real root and  $\theta$ is taken to be  one of these real roots. The other way  to make $\textrm{SRF}^{w}(\theta)$ equal to  zero is when the weights $w$'s have a symmetrical distribution and  $\theta=E_{w}(w)$, i.e., Method II. However, when Method II is used to construct $g^{w}_{n}(\theta)$,  having an  $\textrm{SRF}^{w}(\theta)$ that is   exactly  zero, as it was  already mentioned in Section \ref{Main Results},  will come at the expense of having  confidence regions for $\mu$ whose volumes  do not vanish  (see Section \ref{Length of the Confidence Intervals for mu} and Appendix I).

\subsection*{Edgeworth expansions for  $g_{n}^{w}(\theta)$ in view of  Methods I and II}\label{Edgeworth exapnsions}
We use  Edgeworth expansions to illustrate  the higher accuracy  of the CLT for the randomized pivot  $g_{n}^{w}(\theta)$, as in (\ref{eq 2}),    under Methods I and II, as compared to that of  the classical CLT for the   pivot $t_n$, as in  (\ref{eq 1}).     Edgeworth expansions are used in our reasoning below  since    they provide a direct  link  between the skewness of a pivotal quantity   and the error admitted by its  CLT.

\par
In order to state  the Edgeworth expansion for the sampling distribution  $P_{X,w}\big( g_{n}^{w}(\theta)\leq t  \big)$, for all  $t \in \mathds{R}$,  we first define
\begin{equation}\label{eq 2+1}
Z^{w}_{n}(\theta):=\sum_{i=1}^n (w_i -\theta) (X_i -\mu)\big/\sqrt{n \sigma_{X}^2E_{w}(w_1-\theta)^2}.
\end{equation}
Also, we consider  arbitrary   positive $\epsilon$ and $\epsilon_1$,  and we let $\epsilon_2>0$ be so that $\Phi(t+\epsilon)-\Phi(t)\leq \epsilon_2$, where  $\Phi$ stands for  the standard normal distribution function.

\par
In view  of the above setup, we now write the following approximation.

\begin{eqnarray}
&&- (\frac{\epsilon_1}{\epsilon})^{2} - P_{X} \big(|S^{2}_n -\sigma_{X}^2|>\epsilon_{1}^{2} \big)+ P_{X,w} (Z^{w}_{n}(\theta)\leq t-\epsilon)-\Phi(t-\epsilon)-\epsilon_2 \nonumber\\
&\leq& P_{X,w}(g^{w}_{n} (\theta) \leq t)-\Phi(t)\nonumber \\
&\leq&(\frac{\epsilon_1}{\epsilon})^{2} + P_{X} \big(|S^{2}_n -\sigma_{X}^2|>\epsilon_{1}^{2} \big)+P_{X,w} (Z^{w}_{n}(\theta)\leq t+\epsilon)-\Phi(t+\epsilon)+\epsilon_2.\nonumber\\
&&\label{eq 2'}
\end{eqnarray}

\par
Under the  assumption  $E_{X}|X_1|^3<+\infty$, from   Baum  and Katz  \cite{Baum and Katz}, we conclude  that, as $n\rightarrow +\infty$,
\begin{equation*}
P(|S_{n}^{2}-\sigma_{X}^2|>\epsilon^{2}_1)=O(1/(\sqrt{n}\log^2 n)).
\end{equation*}
By virtue of this result,  we conclude    that replacing $g_{n}^{w}(\theta)$ by $Z_{n}^{w}(\theta)$ produces an error that  approaches zero at the rate $o(1/\sqrt{n})$, as $n \rightarrow +\infty$.

\par
Combining now the preceding  conclusion with (\ref{eq 2'}) and letting  $\varepsilon:=({\epsilon_1}/{\epsilon})^{2}+\epsilon_2$,   we arrive at
\begin{eqnarray}
&&-\varepsilon - o(1/\sqrt{n})+ P_{X,w} (Z^{w}_{n}(\theta)\leq t-\epsilon)-\Phi(t-\epsilon) \nonumber\\
&\leq& P_{X,w}(g^{w}_{n} (\theta) \leq t)-\Phi(t)\nonumber \\
&\leq&\varepsilon + o(1/\sqrt{n})+P_{X,w} (Z^{w}_{n}(\theta)\leq t+\epsilon)-\Phi(t+\epsilon).\label{eq 3}
\end{eqnarray}
The preceding    relation  implies  the asymptotic equivalence of
\begin{equation*}
\big(P_{X,w}(g^{w}_{n}(\theta) \leq t)-\Phi(t)\big) ~~  \textrm{and} ~~  \big( P_{X,w} (Z^{w}_{n}(\theta)\leq t)-\Phi(t) \big)
\end{equation*}
up to an error of order $o(1/\sqrt{n})$. In view of this equivalence and also  recalling that   in  both Methods I and II  the weights have a finite third moment,  we  write  a one-term   Edgeworth expansion for $P_{X,w} (Z^{w}_{n}(\theta)\leq t)$, $t \in \mathds{R}$,    as follows:

\begin{eqnarray}
&& P_{X,w}\big( Z_{n}^{w}(\theta)\leq t  \big)-\Phi(t)\nonumber\\
&=& -(\frac{\phi(t)H_1 (t)}{3!\sqrt{n}} ) \ \Big( \textrm{SRF}^{w}(\theta) \Big)  \   (\frac{E_{X}(X_1-\mu)^3}{\sigma^{3}_{X}})+o(1/\sqrt{n}), \label{eq 4}
\end{eqnarray}
where   $\phi$ is the density  function  of the standard normal distribution and $H_1 (t)= t^2 -1$.

\par
 Under the condition  $E_{X} |X_1|^3 <+\infty$, the following (\ref{eq 5}) and (\ref{eq 6}) are the respective  counterparts  of the approximations  (\ref{eq 3}) and  (\ref{eq 4}) for the classical  $t_n$, as in (\ref{eq 1}), and they read as follows:

\begin{eqnarray}
&&-\varepsilon - o(1/\sqrt{n})+ P_{X} (Z_{n}\leq t-\epsilon)-\Phi(t-\epsilon) \nonumber\\
&\leq& P_{X}(t_{n} \leq t)-\Phi(t)\nonumber \\
&\leq&\varepsilon + o(1/\sqrt{n})+P_{X} (Z_{n}\leq t+\epsilon)-\Phi(t+\epsilon),\nonumber\\
&&\label{eq 5}
\end{eqnarray}
 where
\begin{equation}\label{eq 4+1}
Z_{n}:= \sum_{i=1}^n (X_i -\mu)/\sqrt{n \sigma_{X}^{2}},
\end{equation}
and

\begin{equation}
 P_{X}\big( Z_n\leq t  \big)-\Phi(t)
= -(\frac{\phi(t)H_1 (t)}{3!\sqrt{n}} )    (\frac{E_{X}(X_1-\mu)^3}{\sigma^{3}_{X}})+o(1/\sqrt{n}). \label{eq 6}
\end{equation}

\par
A comparison between  the  expansions (\ref{eq 6}) and     (\ref{eq 4}) shows how  incorporating the  weights $w$'s and their associated  window   $\theta$,  as specified     in Methods I and  II, results in values of $P_{X,w} \big( g_{n}^{w}(\theta) \leq t \big)$ which are closer to the  standard normal distribution $\Phi(t)$ than those of $P_X (t_n \leq t)$. More precisely, under Methods I and II,  having an      $\textrm{SRF}^{w}(\theta)$, such that $|\textrm{SRF}^{w}(\theta)|$    is   small or negligible,   results in smaller or negligible  values   of the  skewness of  $g^{w}_n (\theta)$, as defined  in (\ref{eq 2}). The latter reduction  of the skewness, when $|\textrm{SRF}^{w}(\theta)|$
 is negligible,  by virtue of (\ref{eq 3}) and (\ref{eq 4}), yields   a one-term  Edgeworth expansion for the sampling distribution of  $ g_{n}^{w}(\theta)$  whose magnitude of error is $o(1/\sqrt{n})$ rather than  $O(1/\sqrt{n})$. On the other hand,   in view of  (\ref{eq 5}) and  (\ref{eq 6}),   the rate of convergence of the CLT for the classical    $t_n$, as in (\ref{eq 1}), is of order $O(1/\sqrt{n})$.

\par
In order to further elaborate on the refinement  provided by the skewness reduction  approach provided by  Methods I and II above, we now assume that the data $X$ and the weights $w$ both  have a finite fourth moment.  In addition to the latter assumption, we also assume   that the data  $X$  satisfy Cram\'{e}r's condition that $\limsup_{|t|\rightarrow +\infty} \big| E_{X} (\exp\{ itX_1\}) \big|<1$.
Cram\'{e}r's  condition is required  for    the sampling distributions  $P_{X,w} \big( g_{n}^{w}(\theta) \leq t\big)$ and $P_{X} \big( t_{n} \leq t\big)$ to admit   two-term  Edgeworth expansions.
\par
It is noteworthy that typical examples of distributions for which Cram\'{e}r's condition holds true  are those with a proper density  (cf. Hall  \cite{Hall}).

\par
Once again here,  replacing  $g_{n}^{w}(\theta)$ by $Z_{n}^{w}(\theta)$, as in (\ref{eq 2}) and (\ref{eq 2+1}),  generates the error term $P_{X} (\big|S_{n}^2-\sigma_{X}^2 |>\epsilon^{2}_1)$, where $\epsilon_1$ is an arbitrary small  positive   constant. In view of our moment assumption at this stage,   $E_{X} |X_1|^4 <+\infty$, from  Baum  and Katz  \cite{Baum and Katz} we conclude that, as $n \rightarrow +\infty$

\begin{equation}\label{eq 2'''}
P_{X} (\big|S_{n}^2-\sigma_{X}^2 |>\epsilon^{2}_1)=o(1/n).
\end{equation}
Hence, replacing $g^{w}_{n}(\theta)$  by $Z^{w}_{n}(\theta)$ generates an error of order $o(1/n)$. By virtue of the latter conclusion, an argument similar to the one used to derive (\ref{eq 3}), yields

\begin{eqnarray}
&&-\varepsilon - o(1/n)+ P_{X,w} (Z^{w}_{n}(\theta)\leq t-\epsilon)-\Phi(t-\epsilon) \nonumber\\
&\leq& P_{X,w}(g^{w}_{n} (\theta) \leq t)-\Phi(t)\nonumber \\
&\leq&\varepsilon + o(1/n)+P_{X,w} (Z^{w}_{n}(\theta)\leq t+\epsilon)-\Phi(t+\epsilon). \label{eq 3+1}
\end{eqnarray}

\par
Also, a similar argument  to  (\ref{eq 5})   yields

\begin{eqnarray}
&&-\varepsilon - o(1/n)+ P_{X} (Z_{n}\leq t-\epsilon)-\Phi(t-\epsilon) \nonumber\\
&\leq& P_{X}(t_{n} \leq t)-\Phi(t)\nonumber \\
&\leq&\varepsilon + o(1/n)+P_{X} (Z_{n}\leq t+\epsilon)-\Phi(t+\epsilon), \label{eq 3+2}
\end{eqnarray}
where  $Z_n$ is as defined  in (\ref{eq 4+1}).

\par
The approximation result (\ref{eq 3+1}) implies that $g^{w}_{n}(\theta)$ and  $Z^{w}_{n}(\theta)$ are equivalent up to an error of order $o(1/n)$ and  (\ref{eq 3+2}) yields the same conclusion  for $t_n$ and $Z_n$. By virtue of the latter  two equivalences, we now write two-term Edgeworth expansions for $P_{X,w}(Z^{w}_{n}(\theta)\leq t)$ and $P_{X}(Z_n\leq t)$, $t \in \mathds{R}$,  as follows:

\begin{eqnarray}
&& P_{X,w}\big( Z^{w}_{n}(\theta)\leq t  \big)-\Phi(t)\nonumber\\
&=& -\phi(t)\Big\{\frac{H_1 (t)}{3!\sqrt{n}}  \   \Big(\textrm{SRF}^{w}(\theta)\Big) \   (\frac{E_{X}(X_1-\mu)^3}{\sigma^{3}_{X}})\nonumber\\
&+&\frac{H_2 (t)}{4! n} \{(\frac{E_w (w_1-\theta)^4}{E^{2}_{w} (w_1 -\theta)^2})   (\frac{E_{X}(X_1-\mu)^4}{\sigma^{4}_{X}})-3\}\nonumber\\
&+& \frac{H_3 (t)}{2(3!)^2 n} \ \Big( \textrm{SRF}^{w}(\theta) \Big)^2 \  (\frac{E_{X}(X_1-\mu)^3}{\sigma^{3}_{X}})^2
 \Big\}\nonumber\\
&+&o(1/n), \label{eq 7}
\end{eqnarray}
where  $H_1(t)$ is  as in (\ref{eq 4}),   $H_2(t)= t^3-3t$ and  $H_3(t)=t^5-10t^3+15 t$.
\par
As to $Z_n$, it admits   the following   two-term Edgeworth expansion.

\begin{eqnarray}
&& P_{X}\big( Z_{n}\leq t  \big)-\Phi(t)\nonumber\\
&=& -\phi(t)\Big\{\frac{H_1 (t)}{3!\sqrt{n}}     (\frac{E_{X}(X_1-\mu)^3}{\sigma^{3}_{X}})\nonumber\\
&+&\frac{H_2 (t)}{4! n}   (\frac{E_{X}(X_1-\mu)^4}{\sigma^{4}_{X}}-3)
+ \frac{H_3 (t)}{2(3!)^2 n}  (\frac{E_{X}(X_1-\mu)^3}{\sigma^{3}_{X}})^2   \Big\}
+o(1/n).\nonumber\\
&&\label{eq 7+1}
\end{eqnarray}

\par
In view of  (\ref{eq 7}), and also (\ref{eq 3+1}),   when the data and the weights have four moments and the data satisfy Cram\'{e}r's condition, we conclude that for both Methods I and II,  when $|\textrm{SRF}^{w}(\theta)|$ is small,    the CLT for $g_{n}^{w}(\theta)$  becomes  more  accurate. In particular, when  $|\textrm{SRF}^{w}(\theta)|$  is negligible then the CLT for $g_{n}^{w}(\theta)$ is second order  accurate, i.e., of order $O(1/n)$. In contrast, by virtue of  (\ref{eq 7+1}),  and also (\ref{eq 3+2}), one can readily see that, under the same conditions for the data,  the CLT for $t_n$ is  only  first order accurate, i.e., of order $O(1/\sqrt{n})$.

\section{Confidence intervals for $\mu$ based on $g^{w}_{n}(\theta)$}\label{Length of the Confidence Intervals for mu}
In view of Methods I and II, we are now to put the refinement  provided by the randomized  pivots  $g^{w}_{n}(\theta)$, as in (\ref{eq 2}),  to  use by constructing more accurate confidence intervals for  the population mean $\mu$, in the case of univariate data. In this section we also study the  length of these confidence intervals.

\par

The use of $g^{w}_{n}(\theta)$ as a pivot  results in asymptotic  $100(1-\alpha)\%$, $0<\alpha<1$, size confidence intervals for $\mu$ of the  form:

\begin{equation}\label{eq 8}
\mathcal{C}^{w}(\theta)= \big[\min\{M_n,N_n\}, \max\{M_n,N_n\} \big],
\end{equation}
where
\begin{eqnarray*}
M_n&=& \frac{-z_{1-\alpha/2} S_n \sqrt{n E_{w}(w_1-\theta)^2}-\sum_{i=1}^n (w_i -\theta) X_i}{-\sum_{i=1}^n (w_i -\theta)},\\
N_n&=&\frac{z_{1-\alpha/2} S_n \sqrt{n E_{w}(w_1-\theta)^2}-\sum_{i=1}^n (w_i -\theta) X_i}{-\sum_{i=1}^n (w_i -\theta) },
\end{eqnarray*}
and $z_{1-\alpha/2}$ is the $100(1-\alpha/2)$th percentile of the standard normal distribution.

\par
We now examine the length of $\mathcal{C}^{w}(\theta)$   which  is

\begin{equation}\label{eq 9}
Length(\mathcal{C}^{w}(\theta)):= \frac{2 z_{1-\alpha/2} S_n}{ \big| \sum_{i=1}^n (w_i -\theta) \big| \big/\sqrt{n E_{w} (w_1 -\theta)^2}   }.
\end{equation}

It is  easy to see  that, for $\mathcal{C}^{w}(\theta)$ when it is  constructed by the means of Method I, since $\theta\neq E_{w} (w_1)$,  as $n \rightarrow +\infty$,  we have

\begin{equation}\label{eq 9'}
Length(\mathcal{C}^{w}(\theta))=o_{P_{X,w}}(1).
\end{equation}
In other words, choosing the weights and their associated window constants   in accordance with   Method I, to create the randomized pivot $g_{n}^{w}(\theta)$, as in (\ref{eq 2}),   results in confidence intervals for $\mu$ whose lengths   approach zero, as the sample size increases.

\par
On the other hand, in  the scenario of   Method II we have    $\theta=E_w (w_1)$. The latter choice of $\theta$ implies that, as $n \rightarrow +\infty$,   for all $b \in \mathds{R}$

\begin{equation}\nonumber
P_{w}\big( \sum_{i=1}^n (w_i -\theta)\big/\sqrt{n E_{w} (w_1 -\theta)^2} \leq b  \big)\rightarrow \Phi(b).
\end{equation}
The preceding CLT for the weights, in view of (\ref{eq 9}),  implies that, as $n\rightarrow +\infty$

\begin{equation}\label{eq 10}
P_{X,w} \big(  Length(\mathcal{C}^{w}(\theta))   \leq \ell \big) \rightarrow P(2 \sigma_X  z_{1-\alpha/2}\big/{|Z|} \leq \ell),
\end{equation}
where $\ell \in  \mathds{R}$ and  $Z$ is a standard normal random variable.

\begin{remark}\label{Page 14}
In view of (\ref{eq 10}),  the length of a confidence interval based on the pivot $g_{n}^{w}(\theta)$,  when it is  constructed in accordance with Method II,  converges    in distribution to   a  scaled inverse of a  folded standard normal random variable rather than shrinking, while,  as it was seen  in Section \ref{Error of Convergence}, this method  results in CLTs for $g_{n}^{w}(\theta)$ that, under appropriate conditions, are second order   accurate   (cf. (\ref{eq 3+1}), (\ref{eq 7}) and Table 3), recalling  that in  Method II, $\textrm{SRF}^{w}(\theta)=0$.
\end{remark}

\subsection{Numerical examples for Methods I and II }\label{Numerical Examples for Methods I and II}
In this section we present  some  numerical results to illustrate  the    refinement provided by the randomized confidence intervals $\mathcal{C}^{w}(\theta)$, as in (\ref{eq 8}), when the random weights and their associated window constants   are chosen in accordance with   Methods I  and II.  In addition to examining the accuracy in terms  of empirical  probabilities of coverage, here,  we also address the length of the randomized confidence intervals $\mathcal{C}^{w}(\theta)$.

\par
In our numerical studies in Tables 1-3 below, we generate 1000  randomized confidence intervals as in (\ref{eq 8}), with nominal size of $95\%$, using the cut-off points  $\pm z_{1-\alpha/2}=\pm 1.96$ therein, and 1000  classical  $t$-confidence intervals $\mathcal{E}:=\bar{X}_n \pm 1.96 S_{n}/\sqrt{n}$,  based on  the same generated data with the same nominal size and cut-off points.

\par
In Tables 1-3   coverage($\mathcal{C}^{w}(\theta)$) and  length($\mathcal{C}^{w}(\theta)$)   stand, respectively,  for the empirical  probabilities   of coverage and the empirical lengths  of the  generated confidence intervals $\mathcal{C}^{w}(\theta)$. Also,   coverage($\mathcal{E}$) and length($\mathcal{E}$) stand, respectively,  for the  empirical probabilities  of coverage  and the empirical lengths of the generated  $t$-confidence intervals $\mathcal{E}$ with nominal size 95\%.

\par
In the following Tables 1-2, under the  scenario   of Method I,  we examine the higher accuracy  provided by  the randomized pivot  $g_{n}^{w}(\theta)$, as in (\ref{eq 2}),    over the classical $t_n$, as in (\ref{eq 1}).

$\vspace{-.5 cm}$
\begin{table}[H]\label{Table 1}
\caption{
$w\  \protect\subst\ \chi^2(7),\ \theta=9.3,\ \textrm{SRF}^{w}(9.3)\approx - 0.662$  and nominal size 95\%
  }
\small
\vspace{-.4 cm}
\begin{center}
\scalebox{0.82}{
\begin{tabular}{c c  cc cc}
\hline
 &  $n$ &   coverage($\mathcal{C}^{w}(9.3)$) &  length($\mathcal{C}^{w}(9.3))$ &  coverage($\mathcal{E}$)& length($\mathcal{E}$)   \\
\hline\hline
\multirow{3}{*}{\small X ${\substack{d\\=}}$ Binomial$(10,0.1)$} &10& 0.933& 5.545& 0.905& 1.153 \\
                                                                 &20& 0.947& 2.308& 0.921& 0.810 \\
                                                                &30&  0.950& 1.522 & 0.935 & 0.672
                                                                 \\
                         \hline
  \multirow{3}{*}{X ${\substack{d\\=}}$ Poisson$(1)$} &10& 0.931& 5.447& 0.908& 1.204  \\
                                                       &20& 0.943& 2.096& 0.928& 0.861   \\
                                                      &30& 0.945& 1.518& 0.933& 0.705\\
                         \hline
 \multirow{3}{*}{X ${\substack{d\\=}}$ Lognormal$(0,1)$}     &10& 0.897 & 8.027& 0.801 & 2.147  \\
                                                             &20& 0.907& 4.829 & 0.855 & 1.608   \\
                                                             &30& 0.930& 2.973& 0.875& 1.343  \\
                         \hline
 \multirow{3}{*}{X ${\substack{d\\=}}$ Exponential$(1)$}   &10&  0.913& 6.753 &  0.873 & 1.144  \\
                                                           &20&  0.933& 2.740&  0.903& 0.839  \\
                                                           &30&  0.940& 1.617& 0.920 & 0.694   \\
\hline
\multirow{3}{*}{X ${\substack{d\\=}}$ $\chi^{2}(1)$}   &10&  0.890& 7.772 & 0.833& 1.552  \\
                                                        &20&   0.917&  3.363 & 0.878 & 1.159  \\
                                                        &30&   0.927& 2.158 & 0.895 & 0.957   \\
\hline
\multirow{3}{*}{X ${\substack{d\\=}}$ Beta$(5,1)$}        &10&  0.926 & 0.834 & 0.894 &  0.167  \\
                                                          &20& 0.940&   0.336 & 0.923 & 0.121 \\
                                                          &30&  0.946 & 0.227 & 0.929 & 0.099   \\
                                    \hline
\end{tabular}

}
\end{center}
\end{table}

$\vspace{.7cm}$

\begin{table}[H]\label{Table 2}
\small
\vspace{-.4 cm}
\caption{$w\ {\protect\substack{d\\=}}$ Bernoulli(1/3), $\theta=0.58$, $\textrm{SRF}^{w}(0.58) \approx -0.7$
 and nominal size  95\%  }
\vspace{-.4 cm}
\begin{center}
\scalebox{0.82}{
\begin{tabular}{ c c  cc cc   }
\hline
  &  $n$ &  coverage($\mathcal{C}^{w}(0.58)$) &  length($\mathcal{C}^{w}(0.58))$ &  coverage($\mathcal{E}$)& length($\mathcal{E}$)  \\
\hline
\hline
\multirow{3}{*}{X ${\substack{d\\=}}$ Binomial$(10,0.1)$} &10&  0.941& 4.446& 0.913& 1.140   \\
                                                          &20&  0.946& 2.452& 0.928& 0.832 \\
                                                          &30&   0.950& 1.841& 0.934& 0.673 \\
                         \hline
\multirow{3}{*}{X ${\substack{d\\=}}$ Poisson$(1)$}     &10&  0.942 & 4.597& 0.905& 1.235   \\
                                                          &20& 0.947& 2.618 & 0.927& 0.861   \\
                                                          &30&  0.949& 1.906& 0.929& 0.708\\
                         \hline
 \multirow{3}{*}{X ${\substack{d\\=}}$ Lognormal$(0,1)$}     &10& 0.897 & 8.227& 0.808& 2.118  \\
                                                             &20&  0.921 &4.859 &0.849& 1.604    \\
                                                             &30&  0.932& 3.730 &0.870& 1.369 \\
                         \hline
 \multirow{3}{*}{X ${\substack{d\\=}}$ Exponential$(1)$}   &10&  0.928& 4.415  & 0.868 & 1.142   \\
                                                           &20&   0.938 & 2.552 & 0.904 & 0.840   \\
                                                           &30&  0.945 & 1.882 & 0.914 & 0.696   \\

\hline
\multirow{3}{*}{X ${\substack{d\\=}}$ $\chi^{2}(1)$}   &10&  0.909 & 5.993& 0.836& 1.562  \\
                                                        &20&   0.926& 3.464&  0.876& 1.150  \\
                                                        &30&  0.937& 2.630& 0.900& 0.966  \\

\hline
\multirow{3}{*}{X ${\substack{d\\=}}$ Beta$(5,1)$}        &10&  0.937& 0.661 & 0.895&  0.167  \\
                                                          &20& 0.942& 0.368& 0.923& 0.121  \\
                                                          &30&  0.948& 0.264&  0.927&  0.099   \\
                                    \hline
\end{tabular}
}
\end{center}
\end{table}

\begin{remark}
From  Tables 1 and 2, it is evident that    the randomized pivots  $g_{n}^{w}(\theta)$, as in (\ref{eq 2}), when constructed according to Method I, can significantly outperform $t_n$, as in (\ref{eq 1}),   in terms of accuracy.
\end{remark}

\par
In the following Table 3 we  examine numerically  the performance of $g_{n}^{w}(\theta)$ when it is constructed based on     Method II.

\begin{table}[H] \label{Table 3}
\caption{$w \protect\substack{d\\=}$ Normal$(0,1)$, $\theta=0$, $\textrm{SRF}^{w}(0)=0$  and nominal size $95\%$  }
\vspace{-.4 cm}
\small
\begin{center}
\scalebox{0.8}{
\begin{tabular}{ c c  cc cc   }
\hline
  &  $n$ &  coverage($\mathcal{C}^{w}(0)$) &  length($\mathcal{C}^{w}(0))$ &  coverage($\mathcal{E}$)& length($\mathcal{E}$)  \\
\hline \hline
\multirow{3}{*}{X ${\substack{d\\=}}$ Binomial$(10,0.1)$} &10& 0.963& 18.744&  0.892&  1.170 \\
                                                          &20&  0.954& 14.497&  0.922&  0.824 \\
                                                          &100&  0.949& 17.197&  0.948&  0.372 \\
                         \hline
  \multirow{3}{*}{X ${\substack{d\\=}}$ Poisson$(1)$} &10& 0.951& 28.539&  0.899&  1.245  \\
                                                      &20& 0.948& 22.211&  0.933&  0.874   \\
                                                      &100& 0.954& 29.994&  0.946& 0.391\\
                         \hline

 \multirow{3}{*}{X ${\substack{d\\=}}$ Lognormal$(0,1)$}  &10& 0.957& 24.953&  0.894&  1.217  \\
                                                          &20& 0.951& 49.103&  0.84&  1.724  \\
                                                          &100&  0.947& 41.609&  0.909& 0.822 \\
                         \hline

 \multirow{3}{*}{X ${\substack{d\\=}}$ Exponential$(1)$}   &10&  0.944& 30.549&  0.87&  1.235  \\
                                                           &20&  0.956& 27.110&  0.902&  0.870  \\
                                                           &100&  0.953& 19.068&  0.943&  0.389   \\

\hline
\multirow{3}{*}{X ${\substack{d\\=}}$ $\chi^{2}(1)$}   &10&  0.937& 31.819&  0.844&  1.712  \\
                                                        &20&  0.946& 34.559&  0.865&  1.242  \\
                                                        &100&  0.947& 32.376&  0.924&  0.554   \\

\hline
\multirow{3}{*}{X ${\substack{d\\=}}$ Beta$(5,1)$}        &10&  0.948& 3.098& 0.891& 0.175  \\
                                                          &20&  0.950& 2.984& 0.932& 0.122 \\
                                                          &100& 0.952& 3.254& 0.935& 0.054     \\
                                   \hline
\end{tabular}
}
\end{center}
\end{table}
Note that in Table 3, as the sample size increases, the lengths of the confidence intervals $\mathcal{C}^{w}(0)$, as in (\ref{eq 8}) with $\theta=0$ therein,  that are constructed   based on  Method II, fluctuate   rather than shrink    (see (\ref{eq 10})).

\section{Randomized pivots with higher accuracy using  triangular random weights}\label{Multinomially Weighted Pivots}
In this section we put  the scenario of  Method I into  perspective,  and   extend it to also include triangular  weights. The idea here  is to relate the size of the given sample  to the random weights.

\par
In this section,   we let $w^{(n)}, w_{1}^{(n)},\ldots,w_{n}^{(n)}$ be a \emph{triangular} array of random weights that is  \emph{independent} from the data $X, X_1,\ldots,X_n$. The random weights $w^{(n)}$ here, can  \emph{either} be an i.i.d. array of random variables with $E_{w} |w^{(n)}_{1}|^3<+\infty$, \emph{or} they can have a $\mathcal{M}$ultinomial distribution with size $\mathcal{K}_n$,  i.e.,

\begin{equation}\label{eq 14}
 (w^{(n)}_{1},\ldots,w^{(n)}_{n})\  \substack{d\\=}\  \mathcal{M}ultinomial(\mathcal{K}_n; p_{1,n},\ldots,p_{n,n}),
\end{equation}
where $\mathcal{K}_n=\sum_{i=1}^n w^{(n)}_{i}  \rightarrow +\infty$, as $n\rightarrow +\infty$ and $\sum_{i=1}^n p_{i,n}=1$.
\par
We are now to introduce Method I.1, as a generalization of Method I, that can yield asymptotically, in $n$, SRF's whose absolute values are small or negligible.
\\
\\
\textbf{\emph{Method I.1}}: Let  $w^{(n)}_{1},\ldots,w^{(n)}_{n}$ be  as above. Choose a real valued  constant $\theta^*$  in such a way that for  given    $\delta$, so that $|\delta|$ can be arbitrary small or zero,     \\
 \\
(i) $\theta^{*} \neq  \underset{n\rightarrow +\infty}{\lim} E_{w}(w_{1}^{(n)})$ and \\
(ii) {$\displaystyle{\lim_{n\rightarrow +\infty} \textrm{SRF}^{w^{(n)}}(\theta^*):= \ \lim_{n\rightarrow +\infty}\frac{E_{w}\big(w_{1}^{(n)}-\theta^{*}\big)^3}{(E_{w}\big(w_{1}^{(n)}-\theta^{*})^2 \big)^{3/2}}}= \delta,$}
\\
\\
Moreover, as $n \rightarrow +\infty$, $\theta^*$ should also satisfy the following maximal negligibility
 condition.

\begin{equation}\label{eq 12+1}
\textrm{(iii)}\ \max_{1\leq i \leq n} \big( w_{i}^{(n)}-\theta^{*}  \big)^2\big/ n =o_{P_{w}}(1).~~~~~~~~~~~~~~~~~~~~~~~~~~~~~~~~~~~~~~~~~~~~~~~~
\end{equation}

\par
The counterpart of the pivot $g_{n}^{w}(\theta)$, as in (\ref{eq 2}), in the context  of Method I.1 is  the following $g_{n}^{w^{(n)}}(\theta^{*})$ which  is defined as:

\begin{equation}\label{eq 11}
g_{n}^{w^{(n)}}(\theta^{*}):= \sum_{i=1}^n (w^{(n)}_{i}-\theta^{*})(X_{i}-\mu)\big/\Big(S_n \sqrt{n E_{w}(w^{(n)}_{1}-\theta^{*})^2} \Big).
\end{equation}

\par
We note that one  can, equivalently,  replace $n E_{w}(w^{(n)}_{1}-\theta^{*})^2$, in the denominator of $g_{n}^{w^{(n)}}(\theta^{*})$, by $\sum_{j=1}^n (w_{j}^{(n)}-\theta^{*})^2$.

\begin{remark}\label{Remark 3}
The maximal  negligibility condition (\ref{eq 12+1}) is a sufficient condition  for the following CLT,  for all $t\in \mathbb{R}$.

\begin{equation}\label{eq 13}
P_{X,w}\big(g_{n}^{w^{(n)}}(\theta^{*})\leq t \big)\rightarrow \Phi(t), \ \textrm{as}\ n\rightarrow +\infty.
\end{equation}

The preceding CLT  is valid even  when the random sample   $X_1,\ldots,X_n$    has only two moments provided that (\ref{eq 12+1}) holds true.

\par
The   CLT  in (\ref{eq 13})  is a consequence of   the well known  Lindeberg-Feller CLT in a conditional sense. We  further elaborate on the CLT in (\ref{eq 13}) by noting that, in light  of the dominated convergence theorem,   (\ref{eq 13}) follows from the  following  conditional CLT:
\\
\emph{As $\rightarrow +\infty $,  for all $t \in \mathbb{R}$,  (\ref{eq 12+1}) suffices to have}

\begin{equation*}\label{eq 12+2}
P_{X|w}\big(g_{n}^{w^{(n)}}(\theta^{*})\leq t \big)\rightarrow \Phi(t)\ \textrm{in \ probability}-P_{w},
\end{equation*}
\emph{where  $P_{X|w}$ stands for the conditional probability of $X$ given the  weights $w_{1}^{(n)},\ldots,w_{n}^{(n)}$.}
\end{remark}

\par
It is noteworthy that a typical condition under which (\ref{eq 12+1}) holds true is when the identically  distributed triangular  weights $w^{(n)}$'s, for each $n$,    have a finite $k$th moment, where $k\geq 3$,   and $\lim_{n\rightarrow +\infty} E_{w} |w_{1}^{(n)}-\theta^{*}|^k=c$, for some positive constant $c$. The validity of the latter   claim can  be investigated  by an application of Markov's inequality for
$nP_w\big(|w^{(n)}_1 - \theta|/\sqrt{n} >\epsilon \big)$, where $\epsilon$ is an arbitrary positive number.

\subsection{On the $\mathcal{M}$ultinomial random weights}\label{Multinomially weighted pivots}
We now consider a particular form of the $\mathcal{M}$ultinomial distribution (\ref{eq 14}), in which $\mathcal{K}_n=n$ and $p_{i,n}=1/n$ for $1\leq i \leq n$, i.e.,

\begin{equation}\label{eq 14+1}
 (w^{(n)}_{1},\ldots,w^{(n)}_{n})\  \substack{d\\=}\  \mathcal{M}ultinomial(n; 1/n,\ldots,1/n).
\end{equation}

\par
On taking $\theta^{*}=1.32215$, for example,  in Method I.1,   when the weights are $\mathcal{M}$ultiniomially  distributed as in (\ref{eq 14+1}),   the randomized pivot $g_{n}^{w^{(n)}}(\theta^*)$, as in (\ref{eq 11}), assumes the following specific form:

\begin{equation}\label{eq 15}
g_{n}^{w^{(n)}}({\small{1.32215}})= \sum_{i=1}^n (w^{(n)}_{i}-1.32215)(X_{i}-\mu)\big/\Big(S_n \sqrt{n E_{w}(w^{(n)}_{1}-1.32215)^2} \Big).
\end{equation}
The window constant  $\theta^{*}=1.32215$, in view of Method I.1, when the weights are  $\mathcal{M}$ultinomial as in (\ref{eq 14+1}), was obtained from the following three steps:
\\
Step 1: Obtain the general form of  $\textrm{SRF}^{w^{(n)}}(\theta)$ in this case as follows:

\begin{eqnarray}
&&\textrm{SRF}^{w^{(n)}}(\theta)=\frac{E_{w}\big(w_{1}^{(n)}-\theta\big)^3}{\big(E_{w}(w_{1}^{(n)}-\theta)^2 \big)^{3/2}}\nonumber\\
&&= \frac{-\theta^3 + 3\theta^2 -3 \theta  (n(n-1)/n^2+1)+ n(n-1)(n-2)/n^3 +3n(n-1)/n^2+1}{ \big(\theta^2 -2\theta+n(n-1)/n^2 +1 \big)^{3/2} }.\nonumber\\
&& \label{eq 16}
\end{eqnarray}
\noindent
Step 2: Obtain the limit of $\textrm{SRF}^{w^{(n)}}(\theta)$, that was derived in  Step 1, as follows:

\begin{equation}\nonumber
\lim_{n\rightarrow +\infty} \textrm{SRF}^{w^{(n)}}(\theta)= \big(-\theta^3 + 3\theta^2 -6\theta+5\big)\big/\big(\theta^2-2\theta+2 \big)^{3/2}
\end{equation}

\noindent
Step 3: In light of Step 2,  for  $\theta^{*}=1.32215$,  $\lim_{n \rightarrow+\infty} \textrm{SRF}^{w^{(n)}}(1.32215)$   assumes a  value approximately equal to $\delta=0.0001$ which  is negligible.

\par
We note that the maximal negligibility  condition  (\ref{eq 12+1}) holds  for the $\mathcal{M}$ultinomial weights as in (\ref{eq 14+1}). The latter is true since, in this case,  we have   $\lim_{n\rightarrow +\infty} E_{w} (w_{1}^{(n)}-1.32215)^4=c$, where $c$ is a positive number whose value is not specified here (cf. the paragraph following Remark \ref{Remark 3}). By this, we conclude  that, on taking $\theta^{*}=1.32215$,  all the assumptions in Method I.1 hold true for the $\mathcal{M}$ultinomial weights, as in (\ref{eq 14+1}).

\par
In the present context of  $\mathcal{M}$ultinomial weights, the  $100(1-\alpha)\%$ confidence intervals for $\mu$, based on the pivot $g_{n}^{w^{(n)}}({\small{1.32215}})$, as in (\ref{eq 15}),  follow the general form (\ref{eq 8}). However, the fact   that here we have the constrain   $\sum_{i=1}^n w_{i}^{(n)}=n$, enables us  to specify   (\ref{eq 8}) for $\mu$ in this context, as follows:

\begin{equation}\label{eq 16+1}
\mathcal{C}^{w^{(n)}} (\small{1.32215}):= \frac{ \sum_{i=1}^n (w_{i}^{(n)}- 1.32215)X_i \pm   z_{1-\alpha/2}S_n \sqrt{n E_{w}(w^{(n)}_{1}-1.32215)^2} } {{\tiny{0.32215}} n }.
\end{equation}

\par
$\mathcal{M}$ultinomial random variables, of the form (\ref{eq 14+1}), also  appear in the area of the  weighted bootstrap, also known as the  generalized bootstrap (cf., for example, Arenal-Guti\'{e}rrez  and  Matr\'{a}n   \cite{Arenal-Gutierrez and Matran},  Barbe  and  Bertail  \cite{Barbe  and  Bertail}, Cs\"{o}rg\H{o}  \emph{et al}. \cite{Csorgo et al},
Mason   and Newton  \cite{Mason and Newton} and references therein),   where they represent the count of the  number of times each observation  is selected in a re-sampling with replacement
from a given sample. Motivated by this, somewhat remote, relation   between the bootstrap and  our randomized approach in Method I.1, when the weights are as in (\ref{eq 14+1}), we are now to conduct a numerical comparison between the two methods. After some further elaborations on the weighted bootstrap, we present our  numerical results  in Table 4 below.

\par
To explain the  viewpoint of the weighted bootstrap, we first consider a bootstrap sample $X_{1}^{*},\ldots,X_{n}^{*}$ that is drawn with replacement from the original sample $X_1,\ldots, X_n$. Observe now that for the bootstrap sample  mean $\bar{X}^{*}_n:= \sum_{k=1}^n X_{k}^{*}/n$ we have

\begin{equation}\nonumber
\bar{X}^{*}_n= \sum_{i=1}^n  w_{i}^{(n)} X_{i}/n,
\end{equation}
where, for each $i$, $1\leq i \leq n$,  $w_{i}^{(n)}$ is the count of the number of times the index $i$ of $X_i$ was selected. It is easy to observe that the weights $(w_{1}^{(n)},\ldots,w_{n}^{(n)})$ are $\mathcal{M}$ultinomially distributed, as  in (\ref{eq 14+1}),  and they are independent from the data $X_1,\ldots, X_n$.

\par
To conduct our numerical comparisons,  we consider the    bootstrap $t$-confidence  intervals (cf. Efron  and Tibshirani  \cite{Efron and Tibshirani}) that are  generally known to be efficient of the  second order in probability-$P_X$ (cf., for example, Hall  \cite{Hall}, Shao and Tu \cite{Shao and Tu} and Singh \cite{Singh}). To construct a bootstrap  $t$-confidence interval for the population mean $\mu$,  first a large number, say $B$,  of independent bootstrap samples of size $n$ are  drawn from the original  data.  Let us represent them by $X_{1}^{*}(b),\ldots,X_{n}^{*}(b)$, where $1 \leq b \leq B$. The bootstrap version of $t_n$, as in (\ref{eq 1}),  is computed for each one of  these $B$ bootstrap sub-samples to have  $t_{n}^{*}(1),\ldots,t_{n}^{*}(B)$, where

\begin{eqnarray*}
t_{n}^* &:=& \sqrt{n} \big(  \bar{X}^{*}_n -\bar{X}_n)\big/  S_{n}^*\\
&=&\sum_{i=1}^n (w_{i}^{(n)}-1)X_i \big/\big( \sqrt{n} S_{n}^*\big),
\end{eqnarray*}
   $S_{n}^{*^2}$ is the bootstrap sample variance and $w^{(n)}$'s are as in (\ref{eq 14+1}). These $B$  bootstrap $t$-statistics are then sorted  in  ascending order  to have $t_{n}^{*}[1]\leq \ldots \leq t_{n}^{*}[B]$.
When, for example, $B=1000$, a bootstrap $t$-confidence  interval for $\mu$ with the nominal size   $95\%$ is constructed by   setting:

\begin{equation*}\label{eq 17}
\mathcal{C}^{*}:=t^{*}_{n}[25]\leq t_n \leq t^{*}_{n}[975].
\end{equation*}

\par
For  the same nominal size of 95\%, we are now to compare the performance of the randomized  confidence interval $\mathcal{C}^{w^{(n)}} (\small{1.32215})$, as in (\ref{eq 16+1}),    to that of the  bootstrap $t$-confidence interval $\mathcal{C}^{*}$,  in  Table 4 below.

\par
In Table 4, we generate 1000  confidence intervals  $\mathcal{C}^{w^{(n)}} (\small{1.32215})$. To do so,  we use 1000 replications of the data sets $X_1\ldots,X_n$,  and the $\mathcal{M}$ultinomial weights $(w_{1}^{(n)},\ldots w_{n}^{(n)})$, as in (\ref{eq 14+1}). For each one of  the  generated data sets, based on $B=1000$ bootstrap samples, we also generate  1000  bootstrap $t$-confidence intervals   $\mathcal{C}^{*}$, with nominal size of 95\%.

\par
Similarly to our setups for   Tables 1-3, in Table 4,  we let coverage($\mathcal{C}^{w^{(n)}} (\small{1.32215})$) and length($\mathcal{C}^{w^{(n)}} (\small{1.32215})$) stand for the empirical coverage probabilities  and the empirical lengths of the therein generated  randomized confidence intervals  $\mathcal{C}^{w^{(n)}} (\small{1.32215})$. Also, in Table 4, we let coverage($\mathcal{C}^{*}$) and length($\mathcal{C}^{*}$) stand for the empirical  probabilities of  coverage and the empirical lengths  of the   bootstrap  confidence intervals  $\mathcal{C}^{*}$.

\begin{table}[ht]\small\label{Table 5}
\caption{$w^{(n)}$ are as in (\ref{eq 14+1}), $\theta^*=1.32215$,  $\textrm{SRF}^{w^{(n)}}(1.32215) \approx 10^{-4}$  and nominal size  $95\%$  }
\vspace{-.4 cm}
\begin{center}
\scalebox{0.8}{
\begin{tabular}{ c c  cc cc   }
\hline
  &  $n$ &  {coverage($\mathcal{C}^{w^{(n)}} (1.32215)$)} &  length($\mathcal{C}^{w^{(n)}} (1.32215)$) &  coverage($\mathcal{C}^{*}$)& length($\mathcal{C}^{*}$)  \\
\hline \hline
\multirow{3}{*}{X ${\substack{d\\=}}$ Binomial$(10,0.1)$} &13& 0.943& 3.137&  0.971&  Inf \\
                                                          &20&  0.948& 2.570&  0.943& 0.887\\
                                                          &30&  0.950& 2.147& 0.951& 0.702 \\
                         \hline
 \multirow{3}{*}{X ${\substack{d\\=}}$ Poisson$(1)$} &13& 0.958& 3.290& 0.972&  Inf  \\
                                                      &20& 0.952& 2.712& 0.967& 0.929   \\
                                                       &30& 0.946& 2.248& 0.946& 0.731\\
                         \hline
 \multirow{3}{*}{X ${\substack{d\\=}}$ Lognormal$(0,1)$}     &10&   0.923& 6.940&  0.903& 5.609  \\
                                                             &20& 0.937& 5.115& 0.921& 2.748 \\
                                                             &30&0.946& 4.296&  0.931& 1.828 \\
                         \hline
 \multirow{3}{*}{X ${\substack{d\\=}}$ Exponential$(1)$}   &10& 0.950& 3.451& 0.937& 1.637  \\
                                                           &20& 0.952& 2.626& 0.941& 1.073  \\
                                                           &30& 0.953& 2.225& 0.952& 0.769   \\
\hline
\multirow{3}{*}{X ${\substack{d\\=}}$ $\chi^{2}(1)$}   &10& 0.925& 4.688& 0.938& 3.222    \\
                                                        &20& 0.944& 3.684& 0.946& 1.661   \\
                                                        &30&  0.950& 3.060& 0.946& 1.236   \\
\hline
\multirow{3}{*}{X ${\substack{d\\=}}$ Beta$(5,1)$}        &10&  0.943& 0.511& 0.946& 0.238\\
                                                         &20&  0.951& 0.381& 0.943& 0.134 \\
                                                         &30&  0.953& 0.313& 0.956& 0.108 \\
                                    \hline
\end{tabular}
}
\end{center}
\end{table}

The relatively close performance, in terms of accuracy, of the bootstrap $t$-confidence intervals with $B=1000$ bootstrap samples,  and the randomized  pivot $g_{n}^{w^{(n)}}({\small{1.32215}})$, as in (\ref{eq 15}),   in  Table 4 is  interesting. Further refinements to the randomization approach Method I.1 that results in randomized pivots that can   outperform, in terms of accuracy,   Method I.1  are  presented in Method I.2 in Subsection \ref{Fixed sample approach} below.

\par
It is worth noting that the class of  $\mathcal{M}$ultinomial random weights (\ref{eq 14}) is far richer  than  the particular form  (\ref{eq 14+1}). Our focus  on    the latter was mainly the result of  its application in the area of the weighted  bootstrap. Clearly different choices of the size $\mathcal{K}_n$ and/or $p_{i,n}$ in (\ref{eq 14}) yield different  randomizing weights.

\begin{remark}\label{Comparison to the bootstrap}
The use of  the  randomized pivots introduced in this paper to construct confidence intervals for the mean  by no means is  computationally intensive, while  the bootstrap is  a computationally demanding method. Also,  using the  randomization methods discussed in this paper, one does not have to deal with the problem of  how large the number of bootstrap replications $B$, should be. Moreover,  the error reduction methods   introduced in this paper enable  one  to easily trace down  the effect of the randomization on  the length of the confidence intervals in the univariate case, and the volume of the randomized confidence rectangles when  the data are multivariate (cf. (\ref{eq 8}),   Section \ref{Multivariate Pivots} and Appendix I).

It is also worth noting that the randomization framework  allows regulating   the error of an inference by choosing a desired   value for the SRF. This can be done by choosing  the random weights from a  virtually unlimited class,  as characterized  in the above  Method I, Method II, Method I.1  and also  Method I.2 below.
\end{remark}

\subsection{Fixed sample approach to higher accuracy  using triangular random weights }\label{Fixed sample approach}
The approach discussed in Method I.1 considers triangular random weighs, to tie the random weights to the sample size,  and chooses the window constant  $\theta^*$ therein in such a way that it makes the absolute value of the   SRF arbitrarily small, in the limit. Here, we also consider the triangular random weights as described at the beginning of this section and introduce a method to increase the accuracy of the CLT based inferences about the mean  for   fixed sample sizes.
\par
 For each fixed   sample size $n$, the following Method I.2  yields  a  further sharpening of  the  asymptotic    refinement provided by   Method I.1 and it reads as follows:
\\
\\
\textbf{\emph{Method I.2}}: Let the   weights  $w^{(n)}$'s be  as described right above Method I.1. If for a given $\delta$, so that $|\delta|$ can be arbitrary small or zero,  there exist a real value $\theta^{*}$ so that for  the weights   $w^{(n)}$'s, we have
\\
$\textrm{(i)}\ \displaystyle{\theta^{*}\neq \lim_{n\rightarrow +\infty} E_{w}(w_{1}^{(n)})}$,
 \\
$\textrm{(ii)}\ \lim_{n\rightarrow +\infty}\displaystyle{\textrm{SRF}^{w^{(n)}}(\theta^{*}){=\displaystyle{\lim_{n \rightarrow +\infty}\frac{E_{w}\big(w_{1}^{(n)}-\theta^{*}\big)^3}{(E_{w}\big(w_{1}^{(n)}-\theta^{*})^2 \big)^{3/2}}}}= \delta} $ and
\\
$\textrm{(iii)}\
\displaystyle{\max_{1\leq i \leq n} \big( w_{i}^{(n)}-\theta^{*}  \big)^2\big/ n =o_{P_{w}}(1)},
$
\\
then, for each $n$, choose a real valued  constant $\theta_n$   in such a way that it satisfies the following  conditions (iv) and (v).
\\\\
$\textrm{(iv)}\ \displaystyle{\theta_n \neq   E_{w}(w_{1}^{(n)})}$,
\\$\textrm{(v)}\ \displaystyle{ \textrm{SRF}^{w^{(n)}}(\theta_n){:=\displaystyle{\frac{E_{w}\big(w_{1}^{(n)}-\theta_n\big)^3}{(E_{w}\big(w_{1}^{(n)}-\theta_n)^2 \big)^{3/2}}}}= \delta}$.
\\
\\

The viewpoint in Method I.2,  in principle,  requires choosing  different  $\theta_n$ for different sample sizes $n$, for a given $\delta$.  Also, it is not difficult to see that Method I.1 is the asymptotic version of Method I.2.

\par
Under the scenario of Method I.2,  after choosing an appropriate window value  $\theta_n$, for a given $\delta$,  we define the  randomized pivot $g_{n}^{w^{(n)}}(\theta_n)$ as follows:

\begin{equation}\label{eq 12}
g_{n}^{w^{(n)}}(\theta_n):= \sum_{i=1}^n (w^{(n)}_{i}-\theta_n)(X_{i}-\mu)\big/\Big(S_n \sqrt{n E_{w}(w^{(n)}_{1}-\theta_n)^2} \Big).
\end{equation}

\par
The normalizing sequence $n E_{w}(w^{(n)}_{1}-\theta_n)^2$ in the denominator of $g_{n}^{w^{(n)}}(\theta_n)$ can, equivalently, be replaced by $ \sum_{j=1}^n (w^{(n)}_{j}-\theta_n)^2$.

\par
We note  that,  for each  fixed $n$  and given  $\delta$, when $|\delta|$ is small,  Method I.2  and its associated pivots  $g_{n}^{w^{(n)}}(\theta_n)$, as in (\ref{eq 12}),  yield  higher accuracy    than those that result  from the use of   Method I.1 and its associated pivots $g_{n}^{w^{(n)}}(\theta^{*})$, as in (\ref{eq 11}). This is true since, in Method I.2,  the window constants  $\theta_n$ are tailored for each fixed $n$ to make   $\textrm{SRF}^{w^{(n)}}(\theta_n)= \delta$. This is in contrast to the viewpoint of Method I.1 in which the therein defined  skewness reducing  factor  $\textrm{SRF}^{w^{(n)}}(\theta^{*})$  assumes  the  given   value  $\delta$ in the limit.

\par
Despite their   differences in the context of finite samples, both Method I.1 and Method I.2 yield randomized pivots, as in (\ref{eq 11}) and (\ref{eq 12}), that can  outperform their classical counterpart $t_n$, as in (\ref{eq 1}),  in terms of accuracy (see Tables 4 above and also Tables 5 and 6 below).

\par
Under the scenario of Method I.2, the confidence intervals for $\mu$ based on the randomized pivots  $g_{n}^{w^{(n)}}(\theta_n)$,  also admit the general form  (\ref{eq 8}), only with $w^{(n)}$ in place of $w$ and $\theta_n$ in place of $\theta$ therein.   Hence, in the following numerical studies we  denote  them  by $\mathcal{C}^{w^{(n)}} (\theta_n)$.

\par
In order to illustrate the refinement provided by Method I.2,  we consider  random  samples of sizes $n=10$ and $n=20$ from the heavily  skewed  Lognormal(0,1). We also consider  $\mathcal{M}$ultinomially distributed weights  as in (\ref{eq 14+1}).    Choosing the  random weights here to be $\mathcal{M}$ultinomially distributed, as in (\ref{eq 14+1}) is so that  the numerical  results in Tables 5 and 6 below should be comparable to their counterparts  in Table 4 above where  the data have a  Lognormal(0,1) distribution.

\par
On taking $\delta=10^{-4}$ in Method I.2, we saw in Subsection \ref{Multinomially weighted pivots},  that for $\theta^{*}=1.32215$ we have
\\
$\lim_{n\rightarrow +\infty}\displaystyle{\textrm{SRF}^{w^{(n)}}(1.32215){=\displaystyle{\lim_{n \rightarrow +\infty}\frac{E_{w}\big(w_{1}^{(n)}-1.32215\big)^3}{(E_{w}\big(w_{1}^{(n)}-1.32215)^2 \big)^{3/2}}}}\approx 10^{-4}}$ and
 \\
$\max_{1\leq i \leq n} \big( w_{i}^{(n)}-1.32215 \big)^2\big/ n =o_{P_{w}}(1)$.
\\
Recall that for the $\mathcal{M}$ultinomial weights, as in (\ref{eq 14+1}), the general form of $\textrm{SRF}^{w^{(n)}}(\theta)$ was already derived in (\ref{eq 16}).
In view of the latter result, it is easy  to check that  when $n=10$, on taking  $\theta_{10}=1.2601$ we have  $\textrm{SRF}^{w^{(10)}}(1.2601)\approx 10^{-4}$. Also, for $n=20$, taking  $\theta_{20}=1.29129$ yields $\textrm{SRF}^{w^{(20)}}(1.29129)\approx 10^{-4}$.

\par
 Consider now  $\mathcal{C}^{w^{(10)}} (1.2601)$ and $\mathcal{C}^{w^{(20)}} (1.29129)$,  the   confidence intervals for $\mu$ of nominal size  $95\%$ based on Method I.2 and   samples of size $n=10$ and $n=20$,  which    result, respectively,    from setting:

\begin{eqnarray}
&&-1.96\leq g_{10}^{w^{(10)}}(1.2601)= \frac{\sum_{i=1}^{10} (w^{(10)}_{i}-1.2601)(X_{i}-\mu)}{S_{10} \sqrt{10 E_{w}(w^{(10)}_{1}-1.2601)^2} } \leq 1.96, \nonumber\\
&&-1.96\leq g_{20}^{w^{(20)}}(1.29129)= \frac{\sum_{i=1}^{20} (w^{(20)}_{i}-1.29129)(X_{i}-\mu)}{S_{20} \sqrt{20 E_{w}(w^{(20)}_{1}-1.29129)^2} } \leq 1.96. \nonumber
\end{eqnarray}

\par
In the following Tables 5 and 6 we generate  1000 replications of  Lognormal(0,1) data    and   $\mathcal{M}$ultinomial weights, as in (\ref{eq 14+1}), for $n=10$ and $n=20$. We let coverage($\mathcal{C}^{w^{(10)}} (1.2601)$) and coverage($\mathcal{C}^{w^{(20)}} (1.29129)$) stand for the respective empirical  probabilities  of coverage  of $\mathcal{C}^{w^{(10)}} (1.2601)$ and $\mathcal{C}^{w^{(20)}} (1.29129)$. We also let length($\mathcal{C}^{w^{(10)}} (1.2601)$) and length($\mathcal{C}^{w^{(20)}} (1.29129)$) stand for the respective empirical lengths of $\mathcal{C}^{w^{(10)}} (1.2601)$ and $\mathcal{C}^{w^{(20)}} (1.29129)$.

\begin{table}[ht]\label{Table 6}
\caption{$n=10$, $\theta^{(10)}=1.2601$,  $\textrm{SRF}^{w^{(10)}}(1.2601)\approx 10^{-4}$  and nominal size $95\%$  }
\vspace{-.5 cm}
\begin{center}
\begin{tabular}{ c c c  }
\hline
  &   coverage($\mathcal{C}^{w^{(10)}} (1.2601)$) & length($\mathcal{C}^{w^{(10)}} (1.2601))$   \\
\hline \hline
{X ${\substack{d\\=}}$ Lognormal$(0,1)$}     &0.936& 8.23 \\
                      \hline
\end{tabular}
\end{center}
\end{table}

\begin{table}[ht]\label{Table 7}
\caption{$n=20$, $\theta^{(20)}=1.29129$, $\textrm{SRF}^{w^{(20)}}(1.29129)\approx 10^{-4}$  and nominal size $95\%$  }
\vspace{-.5 cm}
\begin{center}
\begin{tabular}{ c c c }
\hline
  &   coverage($\mathcal{C}^{w^{(20)}} (1.29129)$) & length($\mathcal{C}^{w^{(20)}} (1.29129))$   \\
\hline \hline
{X ${\substack{d\\=}}$ Lognormal$(0,1)$}     &0.944 &5.646  \\
                      \hline
\end{tabular}
\end{center}
\end{table}

In comparison between the two  methods Method I.2 and  Method I.1,  the former  can  outperform the latter, for the same weights and the same $\delta$  (see  Tables 5 and 6,  and compare them  to their counterparts in Table 4  in which  Method I.1 and the bootstrap are examined).

\section{Randomized multivariate pivots}\label{Multivariate Pivots}
The skewness reducing  methods introduced in Methods I, II, I.1 and I.2, can be extended to the case when the data are  multidimensional.
\par
In this section, we first restrict our attention to   Method I  and extend it  to address  multivariate  data (see also Remark \ref{Triangular weights for multivariate} below, where Methods I.1 or  I.2 are used to randomize  multivariate data).  We  show how  the randomization technique of Method I can result in more accurate   multivariate CLT's.

\par
To state our results in this section,  we let    $\mathbb{X}_{j}=(X_{1,j},\ldots,X_{p,j} )^{\prime}$, $1\leq j \leq n$,   be  independent copies of a   $p$-variate, $p\geq 1$,  random vector $\mathbb{X}=(X_{1},\ldots,X_{p})^{\prime}$ such that, for some $k\geq 3$, $E_{\mathbb{X}}\|\mathbb{X}_{1}\|^k< +\infty$, where $\| \mathbb{X} \|= \big(\sum_{s=1}^p X^{2}_{s}\big)^{1/2}$. Furthermore, we let $\mathbf{\mu}=E(\mathbb{X})=(\mu_{1},\ldots,\mu_{p})^{\prime}$ and $\mathbf{\Sigma}$  be the  theoretical  mean  and the theoretical  covariance matrix  of the data $\mathbb{X}$. Moreover, for throughout use in this section, we assume that the covariance matrix $\mathbf{\Sigma}$ is positive definite.
\par
We now define the  pivotal quantity $\mathbb{G}_{n}^{(w)}(\theta)$ that  is the multidimensional  version  of  $g_{n}^{w}(\theta)$, as in (\ref{eq 2}), as follows:
\begin{equation}\label{eq 19}
\mathbb{G}_{n}^{(w)}(\theta):= \Big(\frac{\mathbf{S_n}^{-1/2}}{ \sqrt{n E_{w}(w_{1}-\theta)^2}}\Big)\sum_{i=1}^n (w_{i}-\theta )(\mathbb{X}_{i} -\mathbf{\mu} ),
\end{equation}
where the \emph{univariate}  random  weights $w$, that are independent from the data $\mathbb{X}_1,\ldots,\mathbb{X}_n$,   and   the  window  constant $\theta$ are as characterized in  Method I in  Section \ref{Main Results}, and  $\mathbf{S_{n}^{-1/2}}$ is the inverse of  a positive definite square root of the ($p\times p$) sample covariance matrix

\begin{equation}\label{eq 20}
\mathbf{S_n}=  \sum_{j=1}^n (\mathbb{X}_{j}-\bar{\mathbb{X}}_n) (\mathbb{X}_{j}-\bar{\mathbb{X}}_n)^{\prime}\big/(n-1),
\end{equation}
where $\bar{\mathbb{X}}_n=\sum_{j=1}^n \mathbb{X}_{j}$.

\par
The multivariate pivotal quantity $\mathbb{G}_{n}^{(w)}(\theta)$ is a  randomized version of the classical  multivariate $t$-statistic

\begin{equation}\label{eq 21}
\mathbb{T}_n := \mathbf{S_n}^{-1/2}  \sum_{i=1}^n (\mathbb{X}_{i} -\mathbf{\mu} )\big/ \sqrt{n}.
\end{equation}

\begin{remark}
The possibility of lack of invertibility of  $\mathbf{S_n}$ is a minor drawback that can be resolved  by replacing $\mathbf{S_n}$ by an asymptotically equivalent extended versions of it that are invertible, for all $n$. This  idea is due to Sepanski  \cite{Sepanski}, who proposed  replacing   $\mathbf{S_n}$ by     $\mathbf{D_n}$ that can have either one of the following two forms:

\begin{figure}[t]
$\vspace{-1 cm}$
\includegraphics[width=14 cm, height=8.3 cm]{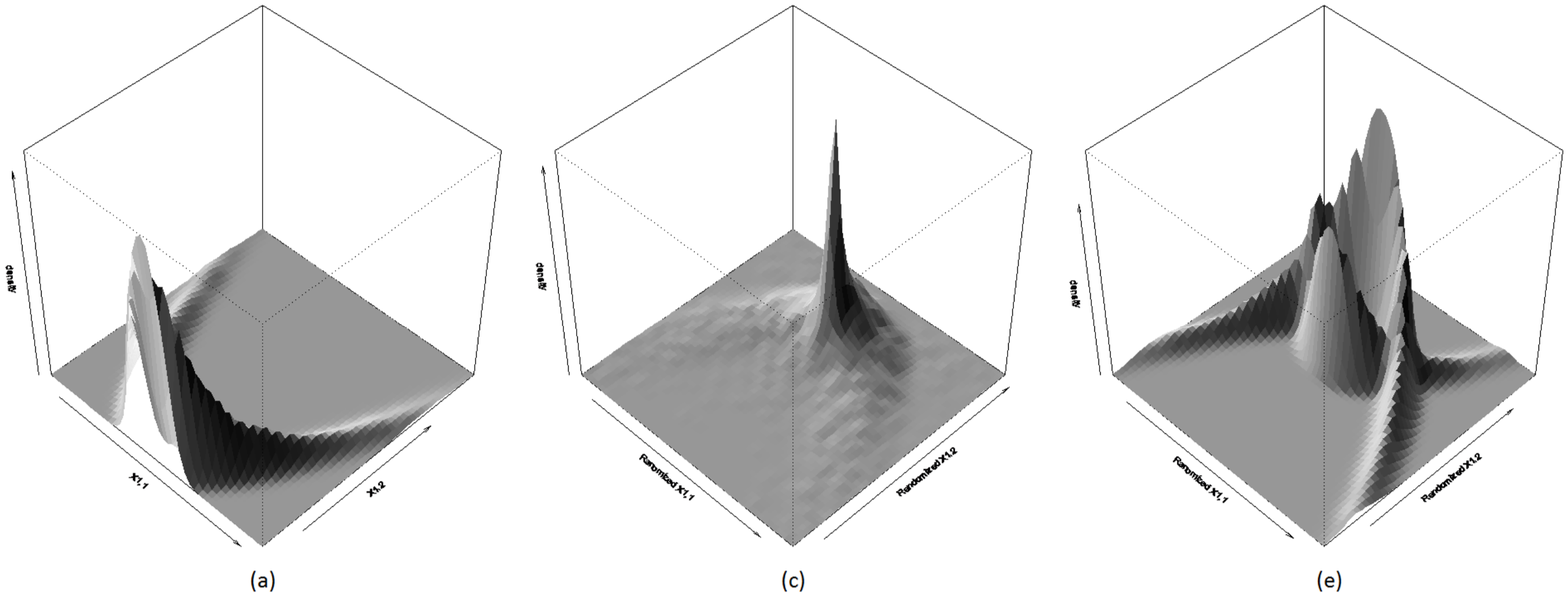}
$\vspace{-2.3 cm}$
\includegraphics[width=14 cm, height=8.3 cm]{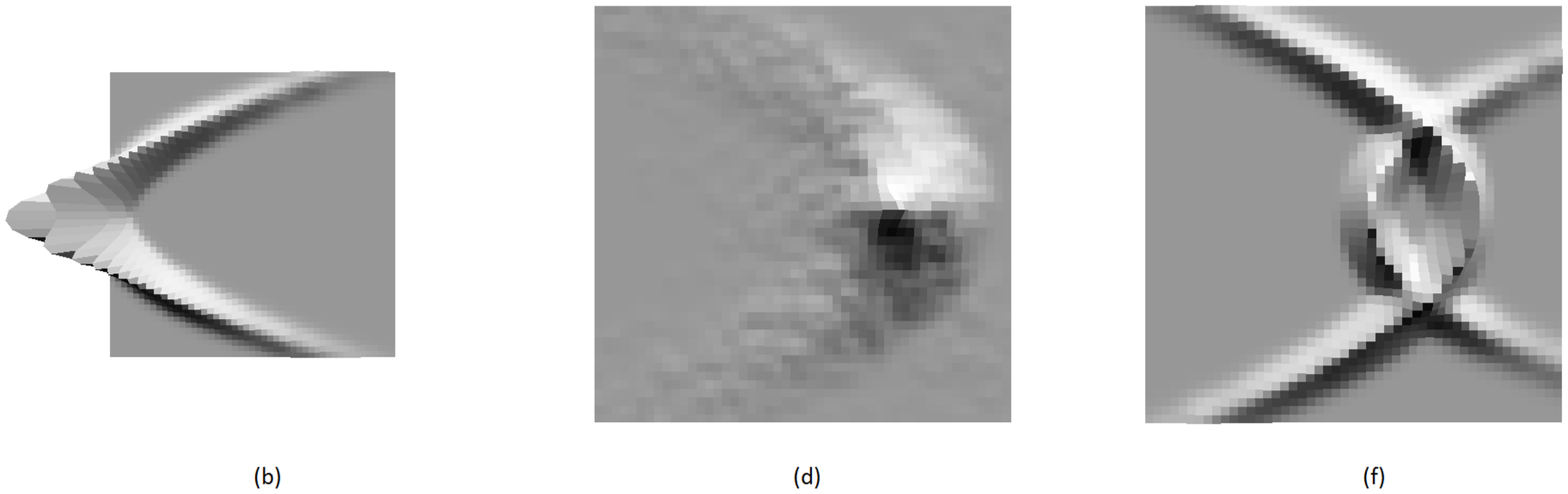}
$\vspace{-2.6 cm}$
\centering
\caption{\footnotesize{ (Illustration of the effect of  Method I on bivariate data)\\
Panels (a) and (b): Two views of the density plot of the original i.i.d. data $(X_{i,1}, X_{i,2})^{\prime}$, $1\leq i \leq 20000$, where ${X_{i,1}\ {\protect\substack{d\\=  }} \ \textrm{Normal}(0,1)}$, $X_{i,2}=X^{2}_{i,1}$, with empirical Mardia's Measure of skewness, cf. Appendix II,  equal to 13.209. Panels (c) and (d):  Two views of the density plot of the randomized data $(w_{i}-9.3)(X_{i,1}, X_{i,2})^{\prime}$, $1\leq i \leq 20000$, where $w_i\ {\protect\substack{d\\=  }}\ \chi^2(7)$, with empirical Mardia's Measure of skewness equal to 3.239. Panels (e) and (f):  Two views of the density plot of the randomized data $(w_{i}-0.58)(X_{i,1}, X_{i,2})^{\prime}$, $1\leq i \leq 20000$, where $w_i\ {\protect\substack{d\\=  }}\ \textrm{Bernoulli}(1/3)$, with empirical Mardia's Measure of skewness equal to 6.216.
}}
\end{figure}

\begin{eqnarray}
\mathbf{D_n} &=& \mathbf{S_n} + l_n \mathbf{I} \label{eq 21+1}\\
\mathbf{D_n} &=& \left\{
                   \begin{array}{ll}
                     \mathbf{S_n}, & \hbox{if  $\mathbf{S_n}\ \textrm{is invertible}$;} \\
                     \mathbf{I}, & \hbox{\textrm{otherwise},}
                   \end{array}
                 \right. \label{eq 21+2}
\end{eqnarray}
where $\mathbf{I}$ is the identity matrix on $\mathbb{R}^p$ and  $l_n $, in (\ref{eq 21+1}),  is a sequence of positive numbers that can approach zero arbitrary fast.
 Hence, when $\mathbf{S_n}$ is not invertible, it can be replaced by either one of the two forms of  $\mathbf{D_n}$, as in  (\ref{eq 21+1}) and (\ref{eq 21+2}),   in  both $\mathbb{G}_{n}^{(w)}(\theta)$ and $\mathbb{T}_{n}$ as in (\ref{eq 19}) and (\ref{eq 21}), respectively.
\end{remark}

\par
In order to show that Method I continues to yield  smaller error for the CLT    for  the randomized multidimensional pivot $\mathbb{G}_{n}^{(w)}(\theta)$,   we first  consider weights and data with a finite fourth moments, i.e., when $E_w |w_1|^4<+\infty $ and $E_{\mathbb{X}} \|\mathbb{X}_{1}\|^4<+\infty$. The refinement provided by  (\ref{eq 19})  under the milder  condition that the data and the weights have a finite third moment  will be discussed later on in this section.

\par
We replace the sample covariance matrix $\mathbf{S}_n$ by the limiting covariance matrix $\mathbf{\Sigma}$. To do so, we adopt the component-wise   convergence in probability, and almost surely, definition  of a sequence of random matrices. More precisely,  we say a sequence of random matrices $\mathbf{A}_n$, $n\geq 1$, converges in probability, or almost surely, to the random matrix $\mathbf{A}$ if each component of $\mathbf{A}_n$ converges in probability, or almost surely, to its counterpart in $\mathbf{A}$. This definition, in turn, enables one  to conclude that replacing the sample covariance matrix  $\mathbf{S}_n$  by the limiting covariance matrix $\mathbf{\Sigma}$, when  $E_{\mathbb{X}}\|\mathbb{X}_1 \|^4<+\infty$,  results in  an error of magnitude $o(1/n)$. The latter statement  is true since both the sample variances and the sample covariances approach their theoretical counterparts at the rate of $o(1/n)$ (see (\ref{eq 2'''})). Consequently,  the multivariate pivot $\mathbb{T}_n$ agrees in distribution with

\begin{equation}\label{eq 21+3}
\mathbb{Z}_n:=\frac{\mathbf{\Sigma}^{-1/2}}{\sqrt{n}} \sum_{i=1}^n (\mathbb{X}_i -\mathbf{\mu})
\end{equation}
up to an error of order $o(1/n)$ where, $\mathbf{\Sigma}^{-1/2}$ is the square root of the inverse of the limiting  covariance matrix $\mathbf{\Sigma}$.


\par
Consider now the standardized data
\begin{equation}\label{eq 22}
\mathbb{Y}_i = (Y_{i,1},\ldots,Y_{i,p})^{\prime}:=
 \mathbf{\Sigma}^{-1/2}
(\mathbb{X}_i-\mathbb{\mu}),\ 1\leq i \leq n,
\end{equation}
and denote  the distribution function of $\mathbb{Z}_n$  by $F_{n,\mathbb{X}} (t_1,\ldots,t_p)$, where $(t_1,\ldots, t_p)\in \mathds{R}^{p}$.    Moreover, let $\Phi(t_1,\ldots,t_p)$ and $\phi(t_1,\ldots,t_p)$ be the respective distribution   and density functions of a   $p$-variate standard normal evaluated at $(t_1,\ldots,t_p)$. Also,  for the ease of  notation,  we define
$$\int_{\prod_{s=1}^p (-\infty,t_s]}\ [...] := \int_{-\infty}^{t_p} \ldots \int_{-\infty}^{t_1} \ [...] \ dt_1\ldots dt_p.$$

\par
Under the assumptions of Theorem 19.2 of Bhattacharya and Rao   \cite{Bhattacharya and Rao}, for all $(t_1,\ldots, t_p)\in \mathds{R}^{p}$,  we have

\begin{eqnarray}
F_{n,\mathbb{X}} (t_1,\ldots,t_p)&=&  \Phi(t_1,\ldots,t_p)\nonumber\\
   &+&   \sum_{j=1}^p  \frac{E(Y^{3}_{1,j})}{\sqrt{n}}  \int_{\prod_{s=1}^p (-\infty,t_s]} -1/6 (-t^{3}_j+3t_j)\phi(t_1,\ldots,t_p)  \nonumber\\
&+& \sum_{1\leq j\neq k \leq p} \frac{E(Y^{2}_{1,j} Y_{1,k})}{\sqrt{n}} \int_{\prod_{s=1}^p (-\infty,t_s]} -1/2 (t^{2}_j t_k +t_k) \phi(t_1,\ldots,t_p)\nonumber\\
&+&  \sum_{\substack{{1\leq j,k,l \leq p}\\{j\neq k, k\neq l, l\neq j}}} \frac{E(Y_{1,j} Y_{1,k} Y_{1,l})}{\sqrt{n}} \int_{\prod_{s=1}^p (-\infty,t_s]}   -t_i t_j t_k \phi(t_1,\ldots,t_p)  \nonumber \\
&+& O(1/n). \label{eq 23}
\end{eqnarray}

\par
As for the randomized i.i.d. data $(w_i-\theta)(\mathbb{X}_i-\mathbf{\mu})=\big( (w_{i}-\theta) (X_{1,i}-\mu_1), \ldots,(w_{i}-\theta) (X_{p,i}-\mu_p) \big)^{\prime}$, let $F_{n,w,\mathbb{X}}(t_1,\ldots,t_p)$ stand for their  distribution function for all $(t_1,\ldots,t_p)\in \mathds{R}^p$.
Consider now the randomized multivariate quantity

\begin{eqnarray}
\mathbb{Z}^{w}_n(\theta)&:=&\frac{\mathbf{\Sigma}_{w,\mathbb{X}}^{-1/2}}{\sqrt{n}} \sum_{i=1}^n (w_i-\theta) (\mathbb{X}_i -\mathbf{\mu})\nonumber\\
&=& \frac{\mathbf{\Sigma}^{-1/2}}{\sqrt{n E_w (w_1-\theta)^2}} \sum_{i=1}^n (w_i-\theta)(\mathbb{X}_i -\mathbf{\mu}),\label{eq 24}
\end{eqnarray}
where $\mathbf{\Sigma}_{w,\mathbb{X}}$ and $\mathbf{\Sigma}$, respectively, are  the covariance matrices  of the randomized data $(w_i-\theta)(\mathbb{X}_i-\mathbf{\mu})$ and the original ones $\mathbb{X}_i$.

\par
An argument similar to the one used to show the asymptotic  equivalence of $\mathbb{T}_n$ and $\mathbb{Z}_n$, as in (\ref{eq 21}) and (\ref{eq 21+3}), enables us   to
also conclude that  $\mathbb{Z}^{w}_n(\theta)$, as in (\ref{eq 24}), is asymptotically equivalent to the randomized pivot $\mathbb{G}^{w}_n (\theta)$, as defined  in (\ref{eq 19}),  at the rate of $o(1/n)$.

\par
By virtue of the above setup we now can write the counterpart of  the  Edgeworth expansion (\ref{eq 23}), under the same conditions on the data $\mathbb{X}_i$,   for the randomized quantity $\mathbb{Z}^{w}_n(\theta)$  as follows:

\begin{eqnarray}
&&F_{n,w,\mathbb{X}} (t_1,\ldots,t_p)=  \Phi(t_1,\ldots,t_p)\nonumber\\
   && +   \sum_{j=1}^p \frac{E_{w,\mathbb{X}}\big( \frac{(w_1-\theta)^3}{E^{3/2}_w(w_1-\theta)^2} Y^{3}_{1,j}\big)}{\sqrt{n}}\int_{\prod_{s=1}^p (-\infty,t_s]} -1/6 (-t^{3}_j+3t_j)\phi(t_1,\ldots,t_p)\nonumber\\
&& + \sum_{1\leq j\neq k \leq p} \frac{E_{w,\mathbb{X}}\big(\frac{(w_1-\theta)^3}{E^{3/2}_w(w_1-\theta)^2}  Y^{2}_{1,j} Y_{1,k}\big)}{\sqrt{n}} \int_{\prod_{s=1}^p (-\infty,t_s]} -1/2 (t^{2}_j t_k +t_k) \phi(t_1,\ldots,t_p)\nonumber\\
&& +  \sum_{\substack{{1\leq j,k,l \leq p}\\{j\neq k, k\neq l, l\neq j}}} \frac{E_{w,\mathbb{X}}(\frac{(w_1-\theta)^3}{E^{3/2}_w(w_1-\theta)^2} Y_{1,j} Y_{1,k} Y_{1,l})}{\sqrt{n}}  \int_{\prod_{s=1}^p (-\infty,t_s]}   -t_i t_j t_k \phi(t_1,\ldots,t_p)  \nonumber
\\
&&+ O(1/n).\nonumber
\end{eqnarray}
Due to independence of the data $\mathbb{X}_i$, and their standardized versions $\mathbb{Y}_i$, as in (\ref{eq 22}),   from the random weights $w_i$, the preceding  Edgewroth expansion is equivalent to the following relation.

\begin{eqnarray}
&& F_{n,w,\mathbb{X}} (t_1,\ldots,t_p) =  \Phi(t_1,\ldots,t_p)\nonumber\\
   && + \  \textrm{SRF}^{w}(\theta)   \Big\{
  \sum_{j=1}^p  \frac{E(Y^{3}_{1,j})}{\sqrt{n}}  \int_{\prod_{s=1}^p (-\infty,t_s]} -1/6 (-t^{3}_j+3t_j)\phi(t_1,\ldots,t_p)  \nonumber\\
&&  + \ \sum_{1\leq j\neq k \leq p} \frac{E(Y^{2}_{1,j} Y_{1,k})}{\sqrt{n}} \int_{\prod_{s=1}^p (-\infty,t_s]} -1/2 (t^{2}_j t_k +t_k) \phi(t_1,\ldots,t_p)\nonumber\\
&& + \ \sum_{\substack{{1\leq j,k,l \leq p}\\{j\neq k, k\neq l, l\neq j}}} \frac{E(Y_{1,j} Y_{1,k} Y_{1,l})}{\sqrt{n}} \int_{\prod_{s=1}^p (-\infty,t_s]}   -t_i t_j t_k \phi(t_1,\ldots,t_p) \Big\}
  \nonumber \\
&& +\  O(1/n).\label{eq 25}
\end{eqnarray}

\par
Denoting now the distribution functions of the multidimensional pivots $\mathbb{T}_n$ and $\mathbb{G}^{w}_n (\theta)$, respectively, by   $Q_{n}(t_1,\ldots,t_p)$ and $Q_{n,w,\mathbb{X}}(t_1,\ldots,t_p)$, from (\ref{eq 23}), as  $n \rightarrow +\infty$,   we conclude that

\begin{equation}\label{eq 26}
Q_{n}(t_1,\ldots,t_p) - \Phi(t_1,\ldots,t_p)=O(1/\sqrt{n}),
\end{equation}
while, under the same conditions on the data $\mathbb{X}$,  the expansion (\ref{eq 25}), as $n \rightarrow +\infty$,  yields

\begin{equation}\label{eq 27}
Q_{n,w,\mathbb{X}}(t_1,\ldots,t_p) - \Phi(t_1,\ldots,t_p)=\big(\textrm{SRF}^{w}(\theta)\big) O(1/\sqrt{n})+O(1/n).
\end{equation}

\par
By virtue of Method I, on  choosing appropriate  random weights $w$ and a window constant $\theta$ to construct $\mathbb{G}_{n}^{(w)}(\theta)$, as in (\ref{eq 19}),   one can achieve CLTs with error rates  up to $O(1/n)$. The optimal rate of $O(1/n)$ is achieved  when $\theta$ is chosen in such a way that $|\textrm{SRF}^{w}(\theta)|$ is negligible. This result is in contrast to the error rate of $O(1/\sqrt{n})$, as in (\ref{eq 26}),  that is the error rate of the CLT for $\mathbb{T}_n$, as in (\ref{eq 21}),  that cannot be improved upon without restricting the class of the distributions of the original data to the symmetrical ones.

\par
Under the milder assumption that the data and the weights have a finite third moment, in view of Theorem 19.2  of Bhattacharya  and Rao   \cite{Bhattacharya and Rao},  using a similar argument as the one used to derive (\ref{eq 27}),  one can conclude the following  statement which is the  counterparts of (\ref{eq 27}) in this context.
\begin{equation}\label{eq 28}
Q_{n,w,\mathbb{X}}(t_1,\ldots,t_p) - \Phi(t_1,\ldots,t_p)=\big(\textrm{SRF}^{w}(\theta)\big) O(1/\sqrt{n})+o(1/\sqrt{n}).
\end{equation}
Once again, a comparison between (\ref{eq 28}) and (\ref{eq 26}) shows that, on assuming that $E\|\mathbb{X}\|^3< +\infty$ and $E_{w}|w_1|^3<+\infty$,   the CLT for the randomized pivot $\mathbb{G}_{n}^{(w)}(\theta)$, as in (\ref{eq 19}), when constructed under the scenario of Method I, will have smaller error as compared to that of $\mathbb{T}_n$, as in (\ref{eq 21}). In particular, when $| \textrm{SRF}^{w}(\theta)|$ is set to be negligible, the CLT for $\mathbb{G}_{n}^{(w)}(\theta)$ is accurate of order $o(1/\sqrt{n})$ rather than $O(1/\sqrt{n})$, as in (\ref{eq 26}), that is the error rate of the CLT for $\mathbb{T}_n$, as in (\ref{eq 21}).

\begin{remark}
The effect of the skewness reduction technique of  Method I on the volume of the simultaneous $p$-dimensional confidence rectangles   for the vector valued mean $\mathbf{\mu}=(\mu_1,\ldots,\mu_p)^{\prime}$, can be addressed  by its effect  on the marginal confidence intervals  for each of the mean components $\mu_i$, $1 \leq i \leq p$. The latter   effect is essentially  the same  as that    discussed in  details in Section \ref{Length of the Confidence Intervals for mu}, in case of univariate data. For details on the effect of randomization on the volume of the randomized confidence rectangles, we refer to  Appendix I.
\end{remark}

\par
The results in Tables 7-12 below are based on  1000 replications of the therein specificated bivariate data and the random weights. As for the cut-off points,   we used $\pm 2.2365$ in Tables 7-12, since $P\big( (Z_1,Z_2) \in [- 2.2365, 2.2365]^2 \big)\approx 0.95$, where $(Z_1, Z_2)$ has a
 standard bivariate normal distribution.

\par
Tables 7-10  are numerical comparisons between the performance of the randomized pivot $\mathbb{G}_{n}^{(w)}(\theta)$, as in (\ref{eq 19}), when constructed according to  Method I, and that of the classical  $\mathbb{T}_n$, as in (\ref{eq 21}).

\begin{table}[H]\small \label{Table 7}
$\vspace{-.1 cm}$
\caption{$w{\protect\substack{d\\=}}$ $\chi^2(7)$, $\theta=9.3$, $\textrm{SRF}^{w}(9.3)\approx - 0.662$  and nominal size $95\%$  }
\vspace{-.4 cm}
\begin{center}
\begin{tabular}{ c c c c   }
\hline
  $\mathbb{X}=(X, X^2)^{\prime}$&  $n$ &  coverage of $\mathbb{G}_{n}^{(w)}(9.3)$ &  coverage of $\mathbb{T}_n$
   \\
\hline \hline
 \multirow{3}{*}{X ${\substack{d\\=}}$ Normal$(0,1)$}      &30& 0.920&  0.863  \\
                                                           &50& 0.933&  0.884 \\
                                                           &100&  0.945&  0.921\\
\hline \hline
\multirow{3}{*}{X ${\substack{d\\=}}$ Exponential$(1)$}  &100& 0.909  & 0.841  \\
                                                         &300&  0.930 & 0.895 \\
                                                         &400& 0.940& 0.903 \\
 \hline
\end{tabular}
\end{center}
\end{table}

\begin{table}[H]\small\label{Table 8}
\vspace{-.5 cm}
\caption{$w$ $\protect\substack{d\\=}$ Bernoulli(1/3), $\theta=0.58$, $\textrm{SRF}^{w}(0.58) \approx -0.7$  and nominal size  $95\%$  }
\vspace{-.4 cm}
\begin{center}
\begin{tabular}{ c c c c   }
\hline
  $\mathbb{X}=(X, X^2)^{\prime}$&  $n$ &  coverage of $\mathbb{G}_{n}^{(w)}(0.58)$ &  coverage of $\mathbb{T}_n$
   \\
\hline \hline
 \multirow{3}{*}{X ${\substack{d\\=}}$ Normal$(0,1)$}      &30&  0.913&  0.845  \\
                                                           &50&  0.936&  0.894 \\
                                                           &100&  0.948&  0.917\\

\hline \hline
\multirow{3}{*}{X ${\substack{d\\=}}$ Exponential$(1)$}  &100&  0.925& 0.835  \\
                                                         &300&  0.939& 0.897 \\
                                                         &400& 0.949& 0.912 \\

 \hline
\end{tabular}
\end{center}
\end{table}

\par
In Tables  9 and 10 below,  the i.i.d. vector valued  data $\mathbb{X}=(\eta_1,\eta_2)^{\prime}$ consist of  the first two terms of the moving average process $\eta_t= \zeta_t+0.2 \zeta_{t-1}$, $t\geq 1$, where $E(\zeta_s)=0$, for $s\geq 0$.

\begin{table}[H]
\vspace{-.2 cm}
\caption{$w{\protect\substack{d\\=}}$ $\chi^2(7)$, $\theta=9.3$, $\textrm{SRF}^{w}(9.3)\approx - 0.662
$  and nominal size $95\%$  }
\vspace{-.4 cm}
\begin{center}
\begin{tabular}{ c c c c   }
\hline
  $\mathbb{X}=(\eta_1, \eta_2)^{\prime}$&  $n$ &  coverage of $\mathbb{G}_{n}^{(w)}(9.3)$ &  coverage of $\mathbb{T}_n$
   \\
\hline \hline
\multirow{2}{*}{$\zeta \ {\protect\substack{d\\=}}$ Normal(0,1)}     &10 & 0.927  & 0.875    \\
                                                               &20 & 0.948 & 0.915  \\

\hline \hline
\multirow{2}{*}{$\zeta\ {\protect\substack{d\\=}}$ Exponential(1)-1}  &30 & 0.932 & 0.885  \\
                                                             & 50 & 0.943  & 0.917  \\

                                    \hline
\end{tabular}
\end{center}
\end{table}

\begin{table}[H]\small\label{Table 11}
\vspace{-.8 cm}
\caption{$w$ $\protect\substack{d\\=}$ Bernoulli(1/3), $\theta=0.58$, $\textrm{SRF}^{w}(0.58) \approx -0.7$  and nominal size $95\%$  }
\vspace{-.4 cm}
\begin{center}
\begin{tabular}{ c c c c   }
\hline
  $\mathbb{X}=(\eta_1, \eta_2)^{\prime}$&  $n$ &  coverage of $\mathbb{G}_{n}^{(w)}(0.58)$ &  coverage of $\mathbb{T}_n$
   \\
\hline \hline
 \multirow{2}{*}{$\zeta \ {\protect\substack{d\\=}}$ Normal(0,1)}     &10 & 0.935  & 0.870    \\
                                                               &20 & 0.950 &  0.912  \\

\hline \hline
\multirow{2}{*}{$\zeta\ {\protect\substack{d\\=}}$ Exponential(1)-1}  &30 & 0.945 & 0.892  \\
                                                             & 50 & 0.950  & 0.917  \\

                                    \hline
\end{tabular}
\end{center}
\end{table}

\begin{remark}\label{Triangular weights for multivariate}
In addition to Method I that was discussed in this section,  in the case of  multivariate data,
  Methods I.1 and I.2,   as stated  in Section \ref{Multinomially Weighted Pivots}     can also result in  significant refinements when they are used to construct the multidimensional  randomized pivot $\mathbb{G}_{n}^{(w)}(\theta)$, as in (\ref{eq 19}),  with $w^{(n)}$ in place of  $w$ therein.
\end{remark}

\par
We demonstrate the validity of Remark \ref{Triangular weights for multivariate}  numerically  in  Tables 11 and 12. To establish the results in Table 12,  we use Method I.1 with the weights having  $\mathcal{M}$ultinomial distribution as in (\ref{eq 14+1}), and $\mathbb{X}=(\eta_1,\eta_2)^{\prime}$ in Table 12, are as in Tables 9 and 10.

\begin{figure}[t]
$\vspace{-2.5 cm}$
\includegraphics[width=14 cm, height=8.3cm]{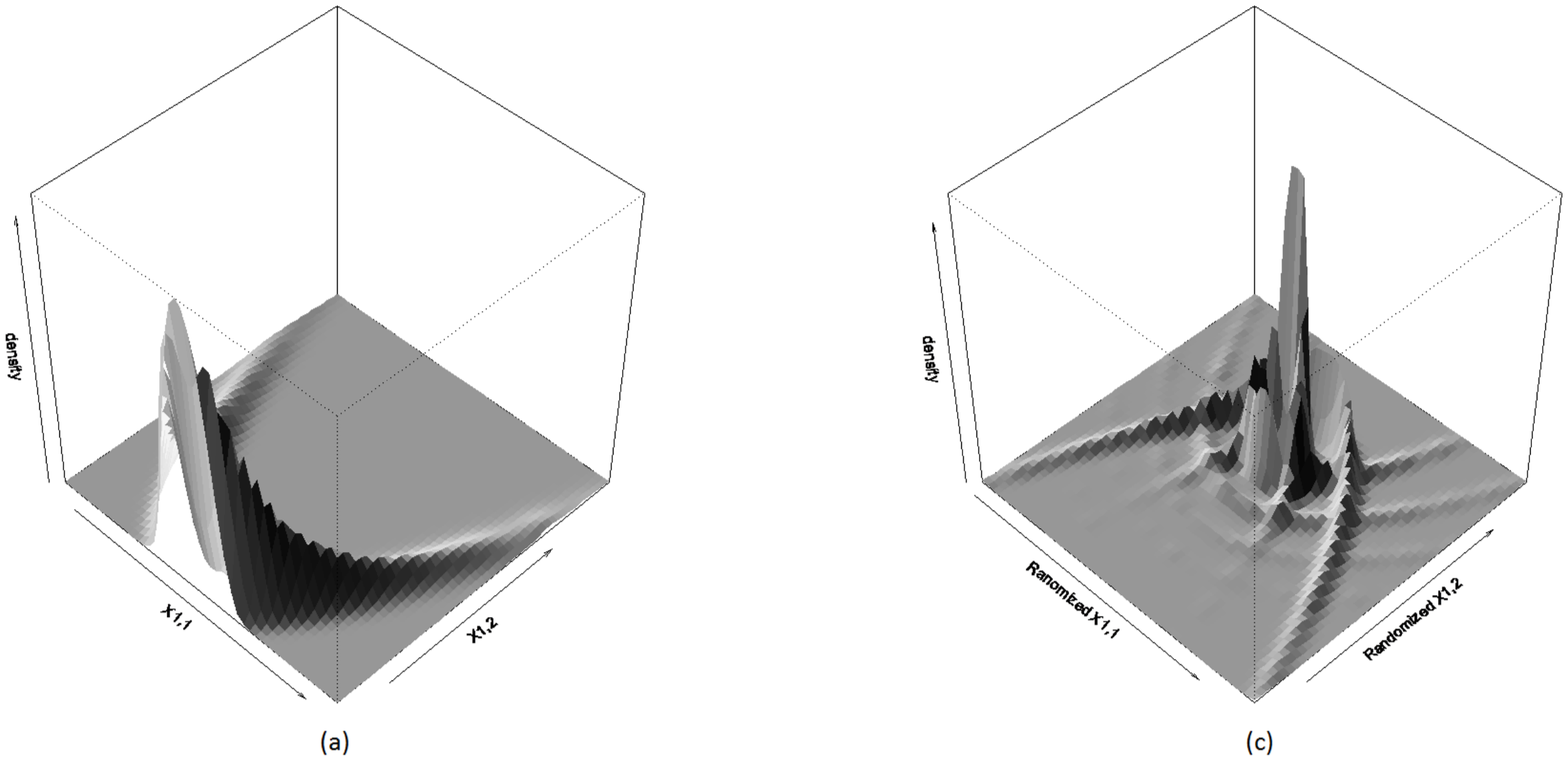}
$\vspace{-.5 cm}$
\includegraphics[width=14 cm, height=8.3cm]{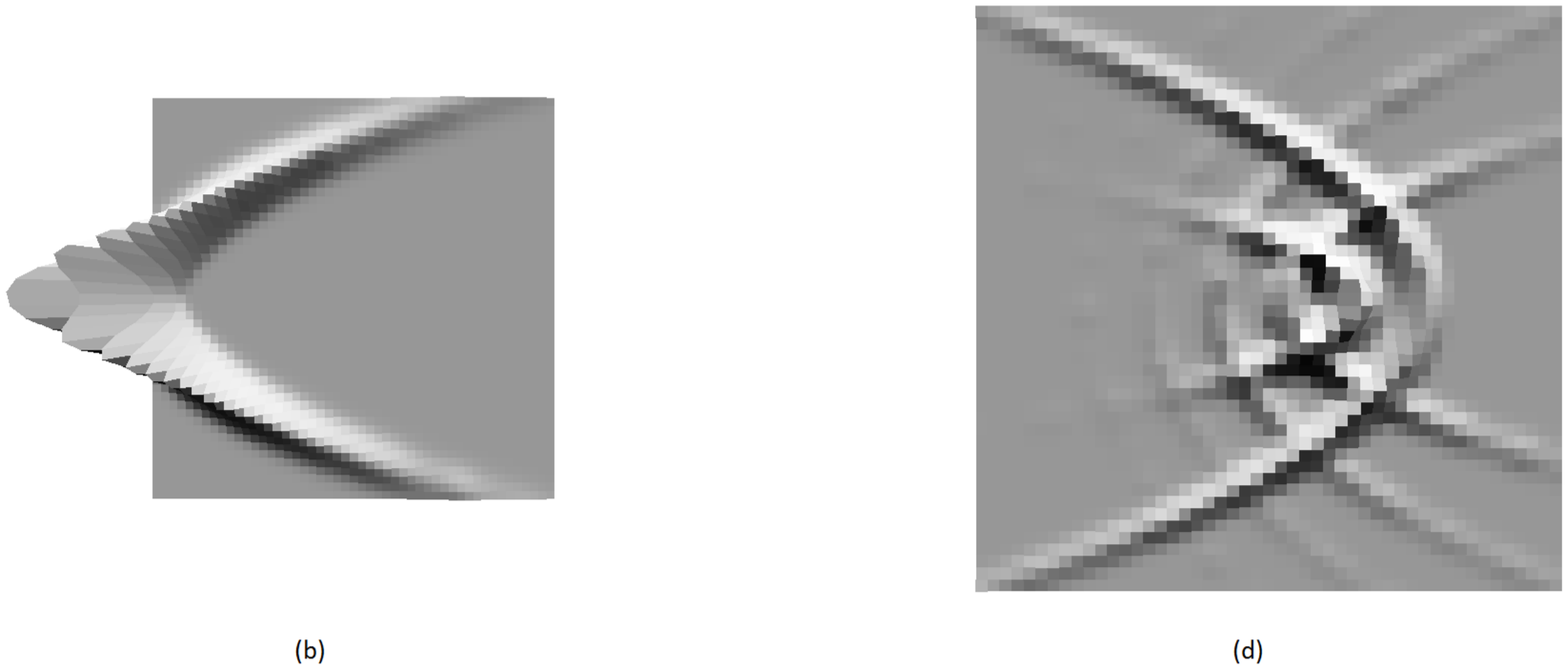}
$\vspace{-2 cm}$
\centering
\caption{\footnotesize{(Illustration of the effect of Method I.1 on bivariate data)\\
Panels (a) and (b): Two views of the density plot of the  data $(X_{i,1}, X_{i,2})^{\prime}$, $1\leq i \leq 20000$, where ${X_{i,1} \ {\protect\substack{d\\= }} \ \textrm{Normal}(0,1)}$ and  $X_{i,2}=X^{2}_{i,1}$. Panels(c) and (d):  Two views of the density plot of the randomized  data $(w^{(n)}_{i}-1.32215)(X_{i,1}, X_{i,2})^{\prime}$, $1\leq i \leq 20000$, where $(w^{(n)}_1,\ldots,w^{(n)}_{20000})\ {\protect\substack{d\\=}} \ \mathfrak{M}\textrm{ultinomial}(20000; 1/20000, \ldots,1/20000)$.
}}
\end{figure}

\vspace{.7 cm}
\begin{table}[H]\small\label{Table 12}
\caption{$w^{(n)}$ are as in (\ref{eq 14+1}), $\theta^*=1.32215$,  $\textrm{SRF}^{w^{(n)}}(1.32215) \approx 10^{-4}$  and nominal size $95\%$  }
\vspace{-.4 cm}
\begin{center}
\begin{tabular}{ c c c c   }
\hline
  $\mathbb{X}=(X, X^2)^{\prime}$&  $n$ &  coverage of $\mathbb{G}_{n}^{(w)}(1.32215)$ &  coverage of $\mathbb{T}_n$
   \\
\hline \hline
 \multirow{3}{*}{X ${\substack{d\\=}}$ Normal$(0,1)$}      &30&   0.933&  0.857  \\
                                                           &50& 0.941&  0.889\\
                                                           &100& 0.950  &  0.917      \\

\hline \hline
\multirow{3}{*}{X ${\substack{d\\=}}$ Exponential$(1)$}  &100& 0.926& 0.860  \\
                                                         &300&  0.940& 0.894 \\
                                                         &400& 0.947  & 0.906   \\

 \hline
\end{tabular}
\end{center}
\end{table}

\begin{table}[H]\small\label{Table 13}
\vspace{-.5 cm}
\caption{$w^{(n)}$ are as in (\ref{eq 14+1}), $\theta^*=1.32215$,  $\textrm{SRF}^{w^{(n)}}(1.32215) \approx 10^{-4}$  and nominal size $95\%$  }
\vspace{-.4 cm}
\begin{center}
\begin{tabular}{ c c c c   }
\hline
  $\mathbb{X}=(\eta_1, \eta_2)^{\prime}$&  $n$ &  coverage of $\mathbb{G}_{n}^{(w)}(1.32215)$ &  coverage of $\mathbb{T}_n$
   \\
\hline \hline
 \multirow{2}{*}{$\zeta \ {\protect\substack{d\\=}}$ Normal(0,1)}     &10 & 0.955  & 0.870    \\
                                                                      &20 & 0.951 &  0.901  \\

\hline \hline
\multirow{2}{*}{$\zeta\ {\protect\substack{d\\=}}$ Exponential(1)-1}  &30 & 0.944 & 0.893  \\
                                                             & 50 & 0.951  & 0.922  \\

                                    \hline
\end{tabular}
\end{center}
\end{table}

\subsection*{Appendix I: Asymptotically exact size randomized  confidence rectangles}
In the case of  multivariate data, the effect of the randomization methods   discussed in Section \ref{Multivariate Pivots}, on the volume of the resulting randomized (hyper) confidence rectangles can be studied by looking at the  marginal confidence intervals for each component of the mean vector. To further elaborate on the idea, for simplicity  we  restrict our attention to two dimensional data as the idea is the same for data with higher  dimensions. Furthermore, here,   we   borrow the notation used in  Section \ref{Multivariate Pivots},  and note that we first consider the randomization approach of Method I. The effect of the other randomization  methods on the volume of the resulting randomized confidence rectangles are   to be addressed later on.

\par
Consider the i.i.d. bivariate  data $\mathbb{X}_j=(X_{1,j},X_{2,j})^{\prime}$, $1\leq j \leq n$, with mean $\mathbf{\mu}=(\mu_1,\mu_2)^{\prime}$.  Furthermore, for  ease of  notation, let $\mathbf{S}^{-1/2}_n=:\begin{bmatrix}
    a_n       & b_n \\
    b_n       & c_n
\end{bmatrix}$, where $\mathbf{S}_n$, as defined in  (\ref{eq 20})  with $p=2$, is the sample covariance matrix.

\par
The classical  $100(1-\alpha)\%$  confidence rectangle for $\mathbf{\mu}=(\mu_1,\mu_2)^{\prime}$ based on the pivot $\mathbb{T}_n$, as in (\ref{eq 21}),  is as follows:

\begin{equation}\tag{$\star$} \label{classical confidence rectangle}
 \Big[  \sum_{j=1}^n X_{1,j}/n \pm z^{*}_{\alpha} \big( \frac{b_n + c_n}{a_n c_n -b^{2}_n}  \big)/\sqrt{n} \Big] \times \Big[  \sum_{j=1}^n X_{2,j}/n \pm z^{*}_{\alpha} \big( \frac{b_n + a_n}{a_n c_n -b^{2}_n}  \big)/\sqrt{n} \Big],
\end{equation}
where  $P(-z^{*}_{\alpha}\leq Z_1 \leq z^{*}_{\alpha} \cap  -z^{*}_{\alpha}\leq Z_2 \leq z^{*}_{\alpha})=1-\alpha$, and   $(Z_1,Z_2)$ has a standard bivariate normal distribution, i.e.,
 $(Z_1,Z_2) \ {\substack{d\\=} \ \textrm{Normal}\big((0,0)^{\prime}, \mathbf{I}\big)}$.

\par
The area of the confidence rectangle  (\ref{classical confidence rectangle}) is

\begin{equation*}\label{length of classical confidence rectangle}
L_{n,\mathbb{X}}:=(2 z^{*}_{\alpha})^2  \frac{(b_n + c_n)(b_n + a_n)}{n (a_n c_n -b^{2}_n)^2}.
\end{equation*}
Observe now that, as $n \rightarrow +\infty$, under the   moment conditions   assumed for the data  in Section \ref{Multivariate Pivots}, we have  $L_{n,\mathbb{X}}=o_{P_{\mathbb{X}}}(1)$.

\par
The randomized version of the confidence rectangle (\ref{classical confidence rectangle}) for $\mathbb{\mu}=(\mu_1,\mu_2)^{\prime}$, in view of Method I,  and based on the randomized pivot $\mathbb{G}^{(w)}_n(\theta)$, as defined in (\ref{eq 19}),  is of the following form:

\begin{equation}\tag{$\star\star$}\label{randomized confidence rectangle}
 \Big[  \min \{ M_{1,n},N_{1,n} \} , \max \{ M_{1,n},N_{1,n} \} \Big] \times \Big[  \min \{ M_{2,n},N_{2,n} \} , \max\{ M_{2,n},N_{2,n} \Big],
\end{equation}
where
\begin{eqnarray}
M_{1,n}&=& \frac{\sum_{j=1}^n (w_j -\theta) X_{1,j}}{ \sum_{i=1}^n (w_i -\theta)}  - z^{*}_{\alpha} \big( \frac{b_n + c_n}{a_n c_n -b^{2}_n}  \big) (\frac{\sqrt{n E_{w} (w_1 -\theta)^2} }{\sum_{i=1}^n (w_i -\theta)}), \nonumber\\
N_{1,n}&=& \frac{\sum_{j=1}^n (w_j -\theta) X_{1,j}}{ \sum_{i=1}^n (w_i -\theta)}  + z^{*}_{\alpha} \big( \frac{b_n + c_n}{a_n c_n -b^{2}_n}  \big) (\frac{\sqrt{n E_{w} (w_1 -\theta)^2} }{\sum_{i=1}^n (w_i -\theta)}), \nonumber\\
M_{2,n}&=& \frac{\sum_{j=1}^n (w_j -\theta) X_{2,j}} {\sum_{i=1}^n (w_i -\theta) } - z^{*}_{\alpha} \big( \frac{b_n + a_n}{a_n c_n -b^{2}_n}  \big) ( \frac{\sqrt{n E_{w} (w_1 -\theta)^2 }} {\sum_{i=1}^n (w_i -\theta)} ), \nonumber\\
N_{2,n}&=& \frac{\sum_{j=1}^n (w_j -\theta) X_{2,j}} {\sum_{i=1}^n (w_i -\theta) } + z^{*}_{\alpha} \big( \frac{b_n + a_n}{a_n c_n -b^{2}_n}  \big) ( \frac{\sqrt{n E_{w} (w_1 -\theta)^2 }} {\sum_{i=1}^n (w_i -\theta)} ). \nonumber
\end{eqnarray}

\par
The area  of the randomized confidence rectangle (\ref{randomized confidence rectangle}) has the following form:

\begin{equation*}\label{length of randomized confidence rectangle}
L_{n,\mathbb{X},w}(\theta):=(2 z^{*}_{\alpha})^2  \big( \frac{(b_n + c_n)(b_n + a_n)}{ (a_n c_n -b^{2}_n)^2} \big)  \big(  \frac{1}{\sum_{i=1}^n (w_i - \theta)\big/  \sqrt{n E_{w} (w_1-\theta)^2}   }\big)^2.
\end{equation*}
Hence, similarly to the  univariate case, in case of multidimensional data,  under the  conditions  of Section \ref{Multivariate Pivots}, in view of Method I,  as $n \rightarrow +\infty$,  we have  $L_{n,\mathbb{X},w}(\theta)=o_{P_{\mathbb{X},w}}(1)$. In other words,
Method I yields randomized confidence regions for the mean vector, that shrink as the sample size increases.

\par
We  remark  that, in the multivariate case,   Methods I.1 and I.2 also yield randomized  confidence rectangles  of the form (\ref{randomized confidence rectangle}), with the  notation $w$ therein  replaced by $w^{(n)}$,  that shrink as the sample size increases. A similar argument to the one used to derive (\ref{eq 10}) shows that the latter conclusion concerning the shrinkage of the randomized confidence regions, in view of Methods I, I.1 and I.2, does not hold true when the randomized pivot $\mathbb{G}_{n}^{(w)}(\theta)$ is constructed using Method II.

\subsection*{Appendix II: The effect of Method I  on Mardia's measure of skewness}
A number of definitions for the concept of skewness of  multivariate data can be found in the literature when the  assumption of normality is dropped. Mardia's  characteristics  of skewness  for multivariate data, cf.   Mardia  \cite{Mardia}, is, perhaps,  the most popular in the literature. This   measures of skewness is  valid when the covariance matrix of the distribution is nonsingular.
For further discussions and developments  on Mardia's skewness and kurtosis characteristics,  we refer  to Kollo  \cite{Kollo} and references therein.

\par
Mardia's measure   of skewness  for $p$-variate  distributions    is defined as follows:

\begin{equation}\nonumber
\beta_{\mathbb{X},p}:= E_{\mathbb{X}}\{(\mathbb{X}_1-\mathbf{\mu})^{\prime} \mathbf{\Sigma}^{-1}(\mathbb{X}_2-\mathbf{\mu})\}^{3},
\end{equation}
where $\mathbb{X}_1$ and $\mathbb{X}_2$ are i.i.d. and  $\mathbf{\Sigma}^{-1}$ is the inverse of the invertible covariance matrix $\mathbf{\Sigma}$.

\par
The following  reasoning  shows how small values of $|\textrm{SRF}^{w}(\theta)|$, as in Method I,   result in smaller values for Mardia's measure of skewness  for  the randomized vectors $(w-\theta)(\mathbb{X}-\mathbf{\mu})$ as compared to that of $\mathbb{X}$.

\par
Let $\beta_{\mathbb{X},w,p}$ be Mardia's measure of skewness of the randomized data $(w-\theta)(\mathbb{X}-\mathbf{\mu})$ and write

\begin{eqnarray*}
\beta_{\mathbb{X},w,p}&=& E_{w,\mathbb{X}} \{  \frac{(w_{1}-\theta)(w_{2}-\theta)}{ E_{w}(w_{1}-\theta )^2}     (\mathbb{X}_1 -\mathbf{\mu} )^{\prime} \mathbf{\Sigma}^{-1} (\mathbb{X}_2 -\mathbf{\mu} )    \}^3\\
&=&   \frac{E^{2}_{w}(w_{1}-\theta)^3}{ E_{w}^{3}(w_{1}-\theta )^{2}} \ E_{\mathbb{X}}\{(\mathbb{X}_1-\mathbf{\mu})^{\prime} \mathbf{\Sigma}^{-1}\ (\mathbb{X}_2-\mathbf{\mu})\}^{3}\\
&=& \Big(\textrm{SRF}^{w}(\theta)\Big)^2 \ \beta_{\mathbb{X},p},
\end{eqnarray*}
where $(w_{1}-\theta)\mathbb{X}_1$ and $(w_{2}-\theta)\mathbb{X}_2$ are i.i.d. with respect to the joint distribution $P_{\mathbb{X},w}$.
The preceding relation shows that employing Method I enables one to make Mardia's  characteristic of skewness arbitrarily small.

\end{document}